%% file: sample-acmtog-SIGGRAPH-submission.tex
\documentclass[acmtog,screen,nonacm]{acmart}
\usepackage{booktabs} 

\usepackage[ruled]{algorithm2e} 
\usepackage{bm}
\usepackage{amsthm}
\usepackage{algpseudocode}
\usepackage{booktabs}
\usepackage{xcolor}
\usepackage{ulem}
\usepackage{multirow}
\usepackage{makecell}

\SetAlFnt{\small}
\SetAlCapFnt{\small}
\SetAlCapNameFnt{\small}
\SetAlCapHSkip{0pt}

\usepackage{colortbl}
\usepackage{array}

\newcommand{\mbR}{\mathbf{R}}
\newcommand{\mbK}{\mathbf{K}}
\newcommand{\mbx}{\mathbf{x}}

\newcommand{\mbt}{\mathbf{t}}
\newcommand{\mbv}{\mathbf{v}}
\newcommand{\mbw}{\mathbf{w}}
\newcommand{\mbX}{\mathbf{X}}
\newcommand{\mbY}{\mathbf{Y}}

\newcommand{\mbc}{\mathbf{c}}
\newcommand{\mbz}{\mathbf{z}}
\newcommand{\mbU}{\mathbf{U}}
\newcommand{\mbA}{\mathbf{A}}
\newcommand{\mbB}{\mathbf{B}}
\newcommand{\mbS}{\mathbf{S}}
\newcommand{\mbF}{\mathbf{F}}
\newcommand{\mbG}{\mathbf{G}}

\newcommand{\mbQ}{\mathbf{Q}}
\newcommand{\mbT}{\mathbf{T}}
\newcommand{\mbP}{\mathbf{P}}

\newcommand{\secref}[1]{\S~\ref{sec:#1}}
\newcommand{\tabref}[1]{Tab.~\ref{tab:#1}}
\newcommand{\figref}[1]{Fig.~\ref{fig:#1}}
\acmJournal{TOG}

\begin{document}
	\title{CHOICE: Coordinated Human-Object Interaction in Cluttered Environments for Pick-and-Place Actions}
	
    
	\author{Jintao Lu}
	\orcid{0000-0002-4141-6710}
	\affiliation{%
		 \institution{The University of Hong Kong}
		 \city{Hong Kong}
         \state{Hong Kong}
		 \country{Hong Kong}}
	\email{jtlu@cs.hku.hk}
    \author{He Zhang}
    \orcid{0009-0003-8607-9441}
	\affiliation{%
		 \institution{Tencent, RoboticsX}
		 \city{Shenzhen}
		 \country{China}}
	\email{herbzhang@tencent.com}
    \author{Yuting Ye}
    \orcid{0000-0003-2643-7457}
	\affiliation{
		 \institution{Meta Reality Lab}
		 \city{Seattle}
         \state{Washington}
		 \country{USA}}
	\email{yuting.ye@gmail.com}
    
    \author{Takaaki Shiratori}
    \orcid{0000-0002-1012-415X}
	\affiliation{
		 \institution{Meta Reality Lab}
		 \city{Pittsburgh}
         \state{Pennsylvania}
		 \country{USA}}
	\email{takaaki.shiratori@gmail.com}
    
    \author{Sebastian Starke}
    \orcid{0000-0002-4519-4326}
	\affiliation{
		 \institution{Meta Reality Lab}
		 \city{Seattle}
         \state{Washington}
		 \country{USA}}
	\email{sstarke@fb.com}

    \author{Taku Komura}
    \authornote{Corresponding auther}
    \orcid{0000-0002-2729-5860}
	\affiliation{
		 \institution{The University of Hong Kong}
		 \city{Hong Kong}
         \state{Hong Kong}
		 \country{Hong Kong}}
	\email{taku@cs.hku.hk}

	\begin{abstract}
		
         Animating human-scene interactions such as picking and placing a wide range of objects with different geometries is a challenging task, especially in a cluttered environment where interactions with complex articulated containers are involved.
		The main difficulty lies in the sparsity of the motion data compared to the wide variation of the objects and environments, as well as the poor availability of transition motions between different actions, increasing the complexity of the generalization to arbitrary conditions.   
		To cope with this issue, we develop a system that tackles the interaction synthesis problem as a hierarchical goal-driven task. 
		Firstly, we develop a bimanual scheduler that plans a set of keyframes for simultaneously controlling the two hands to efficiently achieve the pick-and-place task from an abstract goal signal such as the target object selected by the user.  
		Next, we develop a neural implicit planner that generates hand trajectories to guide reaching and leaving motions across diverse object shapes/types and obstacle layouts.
		Finally, we propose a linear dynamic model for our DeepPhase controller that incorporates a Kalman filter to enable smooth transitions in the frequency domain, resulting in a more realistic and effective multi-objective control of the character.
		Our system can synthesize a rich variety of natural pick-and-place movements that adapt to different object geometries, container articulations, and scene layouts.
		
	\end{abstract}

	%
	%

	%
	%

	
	\begin{teaserfigure}
		\centering
		\includegraphics[width=\linewidth]{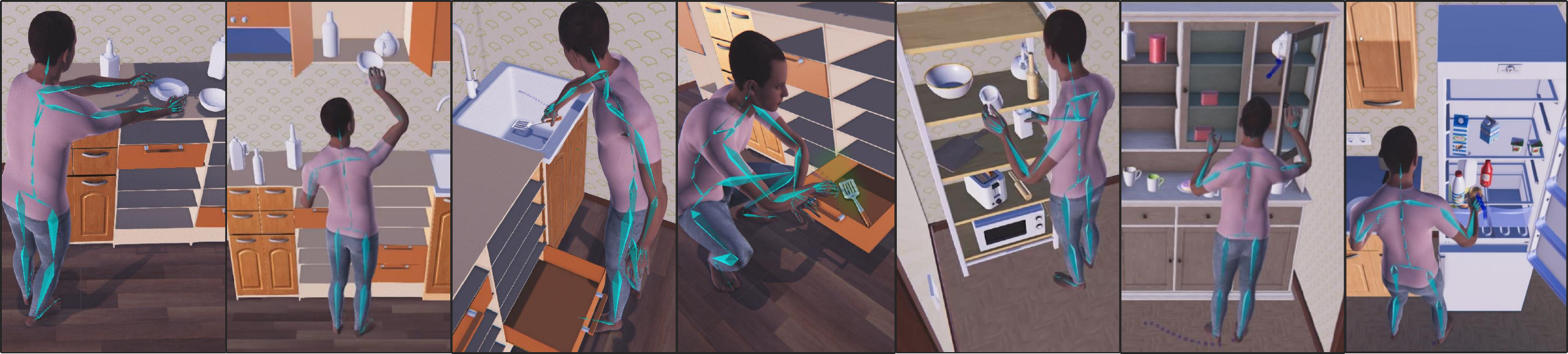}
		\caption{\label{fig:teaser} 
			A selection of multiple interaction tasks in diverse cluttered scenes performed with necessary hand cooperations.
		}
	\end{teaserfigure}

	\maketitle

\input{samplebody-journals}

\end{document}


	\title{Supplement to "CHOICE: Coordinated Human-Object Interaction in Cluttered Environments for Pick-and-Place Actions"}
	
    \author{Jintao Lu}
	\orcid{0000-0002-4141-6710}
	\affiliation{%
		 \institution{The University of Hong Kong}
		 \city{Hong Kong}
         \state{Hong Kong}
		 \country{Hong Kong}}
	\email{jtlu@cs.hku.hk}
    \author{He Zhang}
    \orcid{0009-0003-8607-9441}
	\affiliation{%
		 \institution{Tencent, RoboticsX}
		 \city{Shenzhen}
		 \country{China}}
	\email{herbzhang@tencent.com}
    \author{Yuting Ye}
    \orcid{0000-0003-2643-7457}
	\affiliation{
		 \institution{Meta Reality Lab}
		 \city{Seattle}
         \state{Washington}
		 \country{USA}}
	\email{yuting.ye@gmail.com}
    
    \author{Takaaki Shiratori}
    \orcid{0000-0002-1012-415X}
	\affiliation{
		 \institution{Meta Reality Lab}
		 \city{Pittsburgh}
         \state{Pennsylvania}
		 \country{USA}}
	\email{takaaki.shiratori@gmail.com}
    
    \author{Sebastian Starke}
    \orcid{0000-0002-4519-4326}
	\affiliation{
		 \institution{Meta Reality Lab}
		 \city{Seattle}
         \state{Washington}
		 \country{USA}}
	\email{sstarke@fb.com}

    \author{Taku Komura}
    \authornote{Corresponding auther}
    \orcid{0000-0002-2729-5860}
	\affiliation{
		 \institution{The University of Hong Kong}
		 \city{Hong Kong}
         \state{Hong Kong}
		 \country{Hong Kong}}
	\email{taku@cs.hku.hk}

	\begin{abstract}
		
		This document supplies an extra diagram about the data capture pipeline, the implementation details of our motion controller, and visualizes more cases of performing our trajectory planner under scene generalizations.
		
	\end{abstract}

	%
	%

	%
	%


	\maketitle

	\input{samplebody-journals-Supp}

	\bibliographystyle{ACM-Reference-Format}
\bibliography{sample-bibliography}

%% file: samplebody-journals.tex
\section{Introduction}

Controlling characters to perform pick-and-place motions in cluttered environments with obstacles, such as furniture, has significant applications in video games, virtual reality (VR), and robotics. These environments, often found in household chores, can be surprisingly complex and challenging to automate. The intricacy of the environment escalates due to diverse object shapes, functionalities, and scene layouts with hierarchical levels of objects and varied spatial relations. Planning viable collision-free trajectories for interacting with household objects, like those found in a real-world kitchen with various surface types and tightly packed items, presents an immense challenge. Synthesizing motions for a full-body character that can effectively navigate and interact in such environments is a fundamental yet crucial issue that requires a well-structured approach.

Existing methods that generate human-object interactions in an indoor scene operate in simplified environments.
For example, when grasping and manipulating lightweight objects, the target object is often located in an open space with no obstacles, where the character simply needs to step forward and grasp it~ \cite{wu2022saga,ghosh2023imos}. In reality, however, a milk bottle could be contained in a closed refrigerator that is full of items, or a bowl may be contained in a drawer together with other dishes. A successful interaction with the target object becomes a multi-step process. First, the character needs to prepare an open environment by opening the refrigerator or pulling out the drawer. Next, they need to clear a collision-free path for the hand to reach the target object by moving or even taking out nearby obstacles. Finally, the target object may need to be taken out through a narrow opening without colliding with other objects or the environment.
Throughout the entire process, the person will coordinate all parts of the body for stepping, opening, and grasping tasks according to the layout of the obstacles: 
For example, a person might side-step and open the cabinet with one hand while reaching out for the target object with the other hand.
Although the whole process involves a series of discrete sub-tasks,  the motion is planned such that the entire process is efficient and transitions are smooth.

To tackle these challenges, we propose an integrated system composed of three key modules: a neural implicit trajectory planner that plans collision-free wrist motions, a bimanual scheduler that provides a set of subgoals in terms of keyframes for achieving the goal, and a real-time character controller that synthesizes the motion to reach the planned trajectories and subgoals with smoothness in the DeepPhase~\cite{starke2022deepphase} latent space. Our interaction system approaches the interaction synthesis problem as a hierarchical goal-driven task. Given the complexity of directly dealing with the abstract goal control signals, which only specify the target object and action, we compose three-key-joint goal sequences based on these abstract signals using the bimanual scheduler and the implicit neural trajectory planner. We carefully design the inputs and outputs of each module to progressively transition the abstract and sparse goal towards temporally dense and full-body-level motion.


To construct a valid 6D trajectory of an object to be grasped and picked out/put in a cluttered environment with multiple obstacles, we develop a novel data-driven implicit neural representation from sparse motion capture datasets.   
Our approach is inspired by the recent implicit neural representations of 3D shapes \cite{park2019deepsdf,marschner2023constructive}, scenes \cite{mildenhall2020nerf}, and navigation \cite{li2021learning}.
Their success inspires us to develop a unified representation that describes the spatial-temporal relationship between objects in the environment and the grasping hand. The grasping and placing trajectories from our extensively captured motion dataset could be viewed as an an-isotropic level set propagation process.  Using these data as training samples, 
and SDF of the objects and obstacles as the additional inputs, we construct an implicit time-of-arrival field whose gradient represents the trajectory of the hand to reach the target object at any point in the space.  During inference, our implicit field robustly plans the trajectory functioned by unseen objects and furniture placements.

As our motions involve a series of discrete subtasks by different parts of the body, a naive controller can result in unsmooth and unrealistic coordination and transitions when the interaction goal switches. Moreover, subtle and complex locomotion, such as side-stepping and body twisting, is coupled simultaneously with upper-body interactions. Reproducing such motion realism from demonstrations is challenging.
To cope with this issue, we build our interaction controller upon the DeepPhase model~\cite{starke2022deepphase}, which has demonstrated high motion quality for acyclic motions like dancing. It is well-known that the frequency domain signals can retain the motion details while simplifying the motion representation. We found that the phase provides a compact and efficient low-dimensional feature space to help formulate the full-body motion, serving as a suitable goal feature for control. We then simplify the control law, making it easier to learn after abstracting the motion into our designed DeepPhase state space. Furthermore, we propose 
a linear dynamic model for our DeepPhase controller that incorporates a
Kalman filter to enable smooth state transitions in the frequency domain. This approach further bridges the sparse interaction instructions to continuous-time goals ahead, resulting
in a more realistic and effective 
control of the character while always following the interaction constraints.

To also plan the sequence of motions that are required for reaching the target object, we propose a scheduler for bimanual coordination. By integrating all our proposed methods, we create a comprehensive interaction system.

We demonstrate our system's ability to synthesize realistic pick-and-place interactions in highly complex kitchen environments (see \figref{teaser}). The system can produce subtle movements such as stepping, turning, and manipulation co-temporarily done by two hands in quick response to the user's instructions. Also, it can generalize to diverse scene layouts and shapes.

The key contributions of this work can be summarized as follows:
\begin{itemize}
	
	\item A systematic method for synthesizing long-term interaction in complex environments, hierarchically integrating three innovative sub-modules with significant enhancements individually, verified through quantitative and qualitative experiments;
    
	\item an extensive motion dataset that includes precise full-body manipulation, including the motions of fingers and multiple types of objects, in contextual lifelike indoor scenes, which consists of motions such as taking out/putting in objects from/into containers, manipulation such as pouring, taking off the cap, etc.; 
	
	\item a neural implicit trajectory planner that generates realistic and plausible grasping/placing hand trajectories according to the object type and location, and the layout of the environment;   
	
	\item a DeepPhase-based interaction controller that robustly tracks the goals of two hands and the hip; and 
	
	\item a bimanual scheduler that produces a series of keyframes for the hip and the wrist for pick-and-place tasks of objects within/to be put in various containers, including shelves, drawers, and cabinets. 
	
\end{itemize}

\section{Related Works}

In this section, we first review learning-based motion generation works, especially the RL-based and auto-regressive methods that have the ability for online character-scene interaction.
We then review
research that develops implicit neural representations for the purpose of environmental perceptions and motion planning.

\paragraph{Motion Synthesis of Character-scene Interaction}
The field of motion synthesis has recently seen rapid advancements, utilizing both data-driven and physics-based methodologies~\cite{peng2021amp,peng2022ase,tessler2024maskedmimic}. In the realm of human-scene interaction, kinematics-based humanoid controllers have proven effective in executing interactive motions such as sitting, opening doors, and carrying objects \cite{starke2019neural, zhang2022couch}. By incorporating generative frameworks such as conditional Variational Auto-encoders (cVAE) \cite{hassan2021stochastic}, Diffusion models~\cite{pi2023hierarchical,cen2024generating}, or by further learning a Reinforcement Learning (RL) policy to control the VAE's learned latent space \cite{Zhao:ICCV:2023,luo2023universal}, a more diverse range of interaction behaviors can be synthesized. In the application of physics simulation, diverse interaction motions such as catching and carrying, coupled with locomotion skills, have been realized by distilling a variety of expert demonstrations and RL-based control with visual input \cite{merel2020catch}. More recent progress involves incorporating adversarial training in imitation learning to synthesize realistic interactions \cite{hassan2023synthesizing}. While these approaches yield impressively realistic animations, they are often constrained by the type of objects involved or the complexity of the scene.

Enhanced by the construction \cite{hassan2019resolving} or capture \cite{araujo2023circle} of real-life scenes and motion pairs, human interactions can be synthesized in more clustered and complex scenarios\cite{li2024controllable}. By extending the PROX dataset \cite{hassan2019resolving} with semantic labels and scene segmentation, interaction poses can be conveniently generated through semantic control \cite{Zhao:ECCV:2022}. To compensate for datasets that lack motion-scene pairs, LAMA \cite{lee2023lama} and DIMOS~\cite{zhao2023synthesizing} are introduced to fulfill appropriate motions in the desired scene interaction by incorporating RL. Although these methods can adapt to clustered motions, most interactions are limited to static scenes, lacking details of dynamic objects and finger motions.

To produce fine-grained finger motions for object interaction, a wide range of methods have been explored, including the introduction of canonicalized contact representations \cite{Zheng_2023_CVPR}, grasp fields \cite{karunratanakul2020grasping}, and hand-object spatial representations \cite{zhang2021manipnet}. Leveraging a physics engine, contact sampling \cite{ye2012synthesis}, and DRL-based control \cite{christen2022d, zhang2024artigrasp, hu2024hand} are utilized to synthesize physically plausible manipulations. While these methods focus on isolated hand-object interactions, recent approaches have learned to synthesize full-body motions driven by dynamic object movements \cite{li2023object,li2024controllable}. Incorporating part-wise motion priors (PMP) \cite{bae2023pmp} also allows for the physical assembly of hand and body motions that adapt to complex object movements. More recently, high-quality full-body object manipulation motion datasets have been acquired \cite{taheri2020grab, bhatnagar2022behave, fan2023arctic}, providing opportunities to explore full-body grasping pose synthesis \cite{xu2023interdiff}  and motion control \cite{taheri2022goal, wu2022saga, ghosh2023imos, braun2023physically}. While GOAL \cite{taheri2022goal}, GRIP~\cite{taheri2024grip}, and SAGA \cite{wu2022saga} concentrate on synthesizing full-body approaching and grasping motions, human-object interactions are further enhanced by conditioning on human intention \cite{ghosh2023imos} or physics-based hierarchical control \cite{braun2023physically,luo2024omnigrasp}. However, existing full-body object interaction works are still limited to relatively short motion sequences and monotonous scenes. Drawing from our interaction dataset, we propose a method for synthesizing full-body motion that engages with a variety of scenarios, featuring an array of furniture and everyday objects. Furthermore, we demonstrate that our framework is capable of executing a series of complex tasks involving object and scene interactions, all under interactive user control.

Our control framework design is inspired by a stream of research that has been proposed to discover the periodic motion feature. This includes the periodical feature describing the foot-ground contacting and controlling the character using the phase in the frequency domain \cite{holden2017phase,zhang2018mode,starke2021neural}. DeepPhase \cite{starke2022deepphase} has further proposed the Periodical Auto-Encoder (PAE) that computes a low-dimensional representation of the spatial-temporal body motion. By running a non-linear transformation and FFT, PAE can encode the full-body joint-space motion into several channels of periodic latent vectors, forming a phase manifold in the frequency domain. 

The advantage of the approach is that it can spatio-temporally align a wide range of cyclic and acyclic movements with different hand and leg movements to form a motion manifold in the frequency domain, and then allows detailed auto-regressive control of the body through trajectories projected on the ground.   
Such a characteristic is attractive for our pick-and-place task that requires the combination of complex leg movements, e.g., side-stepping and turning, and co-temporal hand movements, e.g., opening the cupboard on one hand while grasping an object with the other hand, that also requires position-based control for guiding the end effectors to reach the target while avoiding obstacles in the environment.

\paragraph{Implicit Neural Representation for Motion Planning}

The primary concept of using implicit neural representation (INR) methods for 3D geometry involves learning a parametric function via a neural network to represent an objective, such as the signed distance for the query coordinates \cite{park2019deepsdf,marschner2023constructive} or the radiance field for the query view \cite{mildenhall2020nerf}. 
Following its success in shape reconstruction tasks, recent studies have also explored the use of INR-based functions for motion-related activities. 
By constructing the affordance function to determine a potential grasping pose, implicit neural methods have been proposed to learn a field in SE(3) space for the gripper  \cite{simeonov2022neural,manuelli2019kpam,chen2022neural}  or for the finger-hand mesh \cite{karunratanakul2020grasping,yang2022oakink}.
In the realm of navigation,  Li et al.~\shortcite{li2021learning} have utilized an implicit environment field that encodes the maze map and predicts the reaching distance to the target position. Additionally, Camps et al.~\shortcite{camps2022learning} have developed an online deep SDF of the surrounding environment, and NTFields~\cite{ni2022ntfields} navigates the robot through a time field.  Because of its continuity in the latent space, INR naturally extends to feasible grasping for deformed object shapes and navigates the agent using a distance field that fits a different scene.

Most of the current trajectory planning methods are designed for robotic applications, and the trajectory optimization in the canonical space is usually conducted after configuring their potential field~\cite{simeonov2022neural,wang2024implicit}. Similarly, Chen et al.~\shortcite{chen2022neural} apply model predictive control to optimize the objective function that incorporates the evaluation from the grasping field or the distance field, respectively. However, kinematic-based motion generation for virtual characters can often yield more human-like motion patterns. These could include subtle nuances like slightly lifting an object when carrying it out from an upper shelf, or keeping the in-hand object moving with redundant space to potential obstacles while approaching more straightly if the hand is empty. Performing such motion patterns is opposed to merely achieving the objective with the least control cost of the robot arm. 
Therefore, the direct application of trajectory planning will fail to execute some realistic motion patterns that require more control energy. To address this, our proposed implicit neural trajectory planner eliminates the need for an extra stage of adapting trajectory planning algorithms. Instead, it directly learns the realistic wrist trajectory conditioned on the state and environment from the MoCap dataset.

\begin{figure*}[t]
	\includegraphics[width=\linewidth]{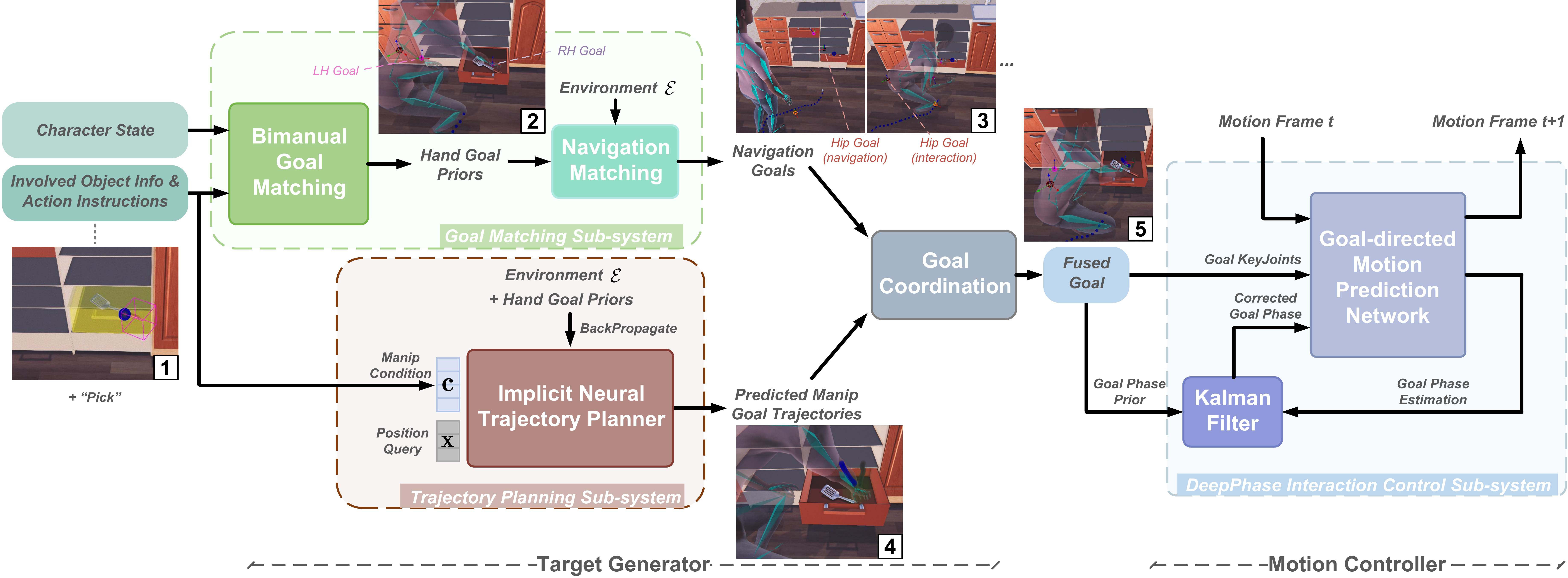}
	\caption{Overview of our CHOICE System: Perceiving the involved objects around the clicking goal object (pic.1) with the action instruction from the keyboard, our system arrange tasks and motions to the empty hands, and match bimanual goals based on the character state and the goal tasks (green block that outputs pic.2). The matched hand goal priors are then re-planned by our trajectory planning sub-system (in dashed brown) to generate a trajectory of manipulation goals that fit the runtime environment (pic.4). After planning a set of navigation goals (pic.3) alongside the manipulation trajectory, our goal coordination arranges the goal for the joints (pic.5) and sets a goal phase prior for the current character. The second major stage of our system is the goal-driven motion controller. It auto-regressively updates the character's motion towards the goal of keyjoints, forming a loop with a Kalman filter, which corrects and sets the goal phase features.}
	\label{fig:pipeline}
\end{figure*}

\section{System Overview}

The structure of our real-time character control system is shown in \figref{pipeline}. We split the interaction synthesis task into three major sub-tasks:
\begin{itemize}
    \item \textbf{hand-object trajectory planning}, managed by the trajectory planning sub-system (dashed brown in ~\figref{pipeline}), which is described in~\secref{INTPPlanner};
    \item \textbf{full-body motion control}, managed by the DeepPhase interaction control sub-system (dashed blue in~\figref{pipeline}), which is described in~\secref{InteractionController}; and
    \item \textbf{bimanual goal scheduling}, managed together by the goal matching sub-system (dashed green in~\figref{pipeline}) and the goal coordination (grey in~\figref{pipeline}), which are described in~\secref{highlevel}. 
    
\end{itemize} When the user clicks the object to pick or place, the path to walk towards the object is first planned, then the series of keyframes for the keyjoints (left/right wrist and hip) based on contacts are scheduled, and finally, collision-free trajectories of the arms to conduct the motion are computed.
The DeepPhase-based controller then synthesizes the motion to follow the scheduled subgoals and planned trajectories in real-time in the phase manifold.
The major mathematical variables involved in these three sub-systems are summarized in~\tabref{math}.

\begin{table}[t]
	\centering
	\caption{Overview of Major Mathematical Variables}
	\label{tab:math}
	\begin{tabular}{|c|c|c|}
		\hline
		\textbf{Sec.} & \textbf{Var.} & \textbf{Description} \\
		\hline
		\multirow{9}{*}{\makecell{4.\\Traj. Plan}} & \( \mathbf{x} \) & Query position \\
		& \( \mathbf{c} \) & Interaction conditions \\
		& \( \mathbf{z} \) & Latent vector of the auto-decoder \\
		& \( D_t \) & Distance field of the target \\
		& \( D_o \) & Distance field of obstacles \\
		& \( D_{toa} \) & Time-of-arrival field of a hand trajectory\\
		& \( \mbx_{prior} \) & Accessiable part of $D_{toa}$ in testing \\
        & \( \tau_{prior} \) & A matched trajectory prior as in~\secref{goalMatching}\\
        & \( \mbx_{s} \) & The predicted trajectory starting position\\
		\hline
		\multirow{6}{*}{\makecell{5.1\\Features}} 
		& \( \mbT_t^{key} \) & 2s trajectory of a key joint pivot on time $t$\\
		& \( \mbP_t \) & Phase feature vector pivoted on time $t$\\
		& \( g \) & Timestamp of the next goal ahead \\
		
		& \( \mbG^{keyJ} \) & Goal transformation of all the keyjoints\\

		& \( \mbT_t^{ego} \) & Ego-centric keyjoint trajectory prediction\\
		& \( \mbT_t^{inv} \) & Goal-centric keyjoint trajectory prediction \\
		
		\hline
		\multirow{6}{*}{\makecell{5.2\\Controller}} 
		
		& \( \mbX_t^p \) & Phase feature pivoted at $t$ as state \\
		
		& \( \mbY_{t} \) & Body pose prediction at $t$ \\
		& \( \mbU \) & Network's control law of phase feature\\
		& \( \mbG^p \) & Goal state of the phase feature\\
		& \( \mbP \) & Covariance matrix of KF estimation\\
		& \( \mbR_{m} \) & Covariance matrix of a measurement\\
		
		\hline
		\multirow{3}{*}{\makecell{6.\\Scheduler}} & \( \bm{p}\) & Character's current pose when matching \\
		& \( \bm{N} \) & Navigation matching feature vector \\
		& \( M \) & Hand-object interaction matching function\\
		\hline
	\end{tabular}
\end{table}

\section{Neural Implicit Trajectory Planner}
\label{sec:INTPPlanner}

In this section, we present the neural implicit trajectory planner designed for the sub-task of generating precise hand grasping or placement of lightweight objects in cluttered scenes with functional furniture (see~\figref{pipeline}, trajectory planning sub-system in dashed brown).
Implicit representation has exhibited remarkable success in learning human faces and body shapes, as well as 3D objects. The motivation behind employing the implicit representation for trajectory planning stems from its exceptional ability to generalize from limited training data. The pick-and-place motion involves intricate object handling, which encompasses various conditions such as diverse geometries, shapes, and articulations of the container, as well as object locations within the container. Additionally, the trajectory planning is influenced by the presence of obstacles of different shapes at various positions along the path, as well as the position and orientation of the character. By leveraging the implicit representations, we can harness their capacity to handle the complexity and variability of these conditions, empowering the controller to devise trajectories that seamlessly adapt to a wide range of scenarios encountered during the pick-and-place task.

We introduce an implicit trajectory planner based on an auto-decoder structure, where the trajectory is inferred by optimizing a time-of-arrival field that represents the trajectory to reach or place the object from all directions. This approach draws inspiration from DeepSDF~\cite{park2019deepsdf}, where an auto-decoder structure is applied to learn spatial implicit fields, providing a compact and continuous latent space that bridges the motion and the scene.

Although DeepSDF can result in over-smoothed surface reconstructions, where high-frequency details are lost, this limitation is not critical for our trajectory planning. To generate smooth, collision-free trajectories, capturing each obstacle's overall shape and position is more important than preserving fine surface details. Moreover, since the feasible region for planning typically maintains a clearance from the obstacle surfaces, and because the palm and fingers usually wrap around an in-hand object, the surface details finer than the fingers are mostly irrelevant to the planning. Indeed, current state-of-the-art trajectory planners~\cite{wang2022geometrically,zhou2019robust,sundaralingam2023curobo} mostly simplify the detailed surface of obstacles with convex primitives.

\begin{figure}
	\includegraphics[width=.9\columnwidth]{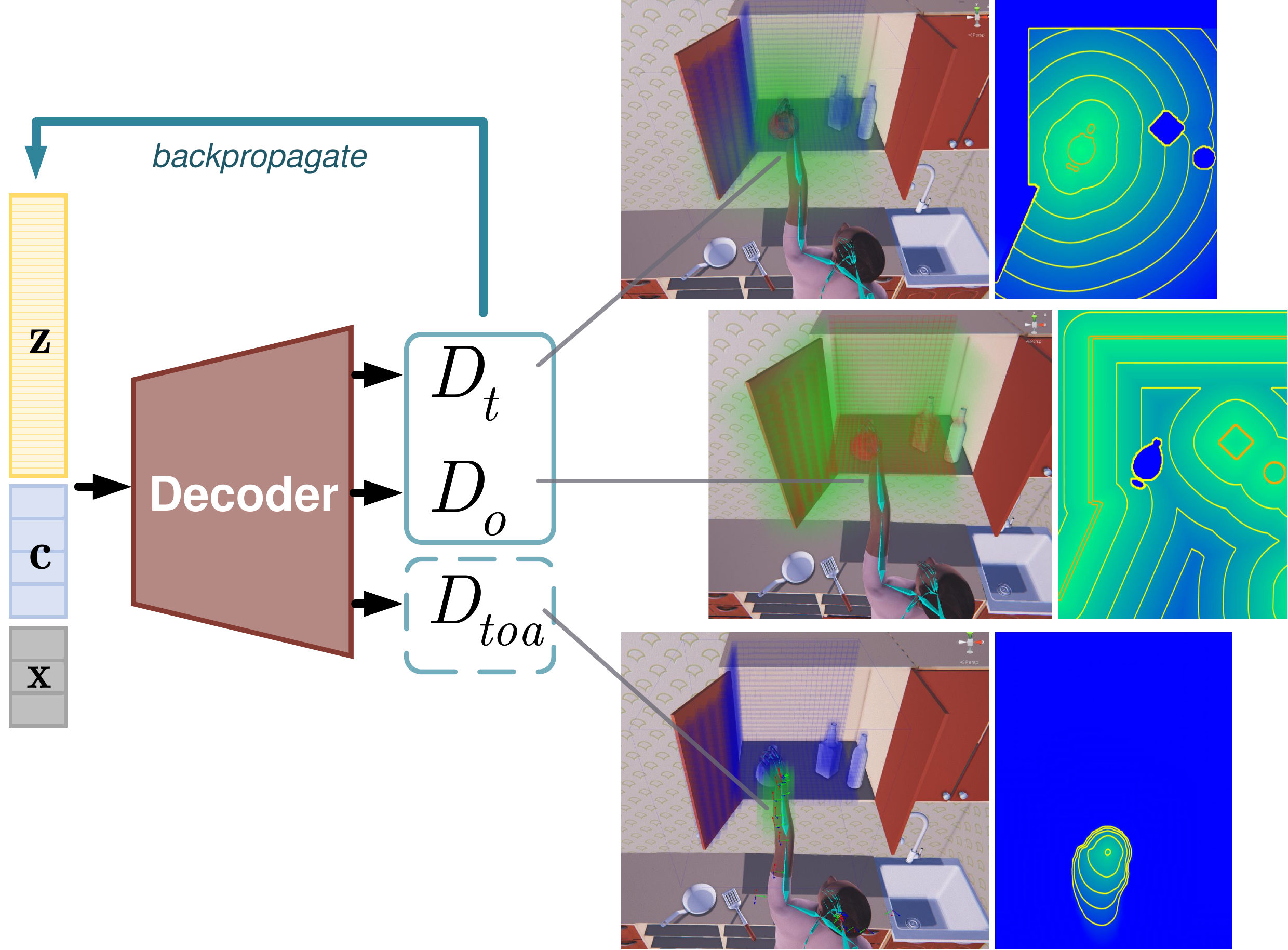}
	\caption{Implicit neural trajectory planner: The scene visualization of the three fields uses the green color to show its value, deeper green represents a lower distance, and the blue color highlights the region of infinity distance. The right-side images give the 2D slices at the teapot height, where the zero-level set was shown in orange curves. Under a test scene, $\mathbf{z}$ was optimized to reconstruct the known part of the output, which was encircled by the blue rectangles. The dashed blue rectangle illustrates the pre-known part of the time-of-arrival field.}
	\label{fig:INTP}
\end{figure}

We first introduce a new field representation that models the hand movements jointly with the geometric information from the environment in~\secref{FieldFeatures}. Then we describe our auto-decoder architecture (see~\figref{INTP}), and how it is trained to synthesize an environment-aware hand trajectory in~\secref{INTP}.

\subsection{Joint Neural Representation of the Hand Trajectory and the Environment}
\label{sec:FieldFeatures}
We describe the 3D space $E$ using features that contribute to the trajectory planning of the hand approaching and leaving. 
The backbone of our system is a 
function $\mathbf{f}$
that maps each local coordinate $\mbx \in R^{3}$ to a feature vector $\mathbf{D}$ as follows:
\begin{equation}
	\mathbf{f}: \mathbb{R}^3 \rightarrow \mathbb{R}^3, \mathbf{f}(\mbx) = \mathbf{D}(\mbx) = (D_{t}(\mbx), D_{o}(\mbx), D_{toa}(\mbx))
\end{equation}
where 
$D_{t}(\mbx)$ is the inverse  unsigned distance of each position $\mbx$ to the target object surface $M_t$, 
$D_{o}(\mbx)$ is similarly the inverse unsigned distance to the set of obstacles 
$M_{O} := \{M_{O_{1}},M_{O_{2}},...,M_{O_{n}}\}$,
and $D_{toa}(\mbx)$ 
represents the time to be travelled by the hand between $\mbx$ for reaching the object: we now describe the details of each term below. 

$D_{t}(\mbx)$ is computed 
using the fast marching method (FMM) \cite{tsitsiklis1995efficient,sethian1996fast}  following the Eikonal equation
\begin{equation}
	\|\nabla d(\mbx)\| = 
	\begin{cases}
		1 & \text{if }  \mbx \ outside\  M_{t} \cup M_{o}\\
		\infty & \text{if } \mbx \ inside\  M_{t} \cup M_{o}
	\end{cases}
\end{equation}
with the boundary condition $d(\mathbf{v}_{i}) = 0$ for all vertices $\mathbf{v}_{i}$ on $M_{t}$. $D_{t}(\mbx)$ then forms a spatially continuous function as follows:
\begin{equation}
	\label{eq:SDFt}
	D_{t}(\mbx) = 
		1/\max(d(\mbx,M_{t}), \epsilon)
\end{equation}
where the reciprocal form makes the field continuous on the boundary near the obstacles, and clipping the unsigned distance makes the field continuous on the boundary near the target object's surface. 

$D_{o}(\mbx)$, the inverse distance field of the obstacle, is similarly formed as follows:
\begin{equation}
	D_{o}(\mbx) =
	1 /  \min \{ \, \max(d(x, M_{O_{i}}), \epsilon)),i = 1,2,...,n \}.
\end{equation}

$D_{toa}(\mbx)$, the time of arrival field, 
represents the 6D trajectory of the hand to reach/leave the mesh surface; the field is learned from each motion capture data and is extrapolated to the volume outside the trajectory, and can further generalize to unseen scenes.
We first describe how we compute $D_{toa}$ for each training data for reaching out for a target object. We generate a potential field $\phi(\mbx)$ such that its value on every point $\mbx$ on the hand trajectory $\tau$ represents the remaining time for the wrist to move from the current point to $\mbx_{goal}$ where the hand first contacts the object. 

We evolve the time-of-arrival of the demonstration trajectory $\tau$ to the 3D space using the Gaussian probability density function (PDF), which decreases the speed value of each grid according to its distance to the trajectory $\tau$. 
Setting the initial value as $\phi(\mbx_{goal}) = 0$, $\phi$ is computed by solving the following Eikonal equation:
\begin{equation}
	\|\nabla \phi(\mbx)\| =
	\begin{cases}
		\frac{1} {||\mathbf{v}(\mbx_n)|| \cdot PDF(d(\mbx,\tau), \sigma)} & \text{if } \mbx \ outside\  M_{t} \cup M_{O} \\
		\infty & \text{if } \mbx \ inside\  M_{t} \cup M_{O}
	\end{cases}
\end{equation}
where $\mbx_n$ is the nearest position on the trajectory $\tau$ from $\mbx$, and the hand speed at $\mbx_n$, $||\mbv(x_n)||$, is set as the peak value of the Gaussian PDF. 
Finally, $D_{toa}$ is computed as an inverse of the time-of-arrival as below:  
\begin{equation}
	D_{toa}(\mbx) = 1/\max(\phi(\mbx), \epsilon))
\end{equation}
which ensures the continuity of the field. 	

Consequently, the three fields formed by $\mathbf{f}$ describe three different but highly related types of representation under the same environment: Querying each position $\mbx$, we can acquire its reaching distance to the goal object, the reaching distance to the nearest obstacle, and an arrival time to the contact frame. 
The three fields are set as the outputs of an implicit neural function $\mathbf{f}_{\theta}$  
which is applied to address the scene understanding and trajectory planning tasks in a combined fashion
as described next in \secref{INTP}.

\subsection{Implicit Neural Trajectory Planning Architecture}
\label{sec:INTP}
In this section, we form an auto-decoder architecture as shown in \figref{INTP} (also corresponds to \figref{pipeline} brown part), which we apply to compute the trajectory to reach or release an object in the environment.  
This is done by optimizing the latent code $\mbz$ and computing the time-of-arrival field $D_{toa}$.
To achieve this task, we first train the auto-decoder using sample fields computed from the motion capture data described in \secref{dataLabeling}.
During runtime, using the trained auto-decoder, we optimize the latent code $\mbz$ such that the loss function that evaluates the L1 loss of the fields and other regularization losses is minimized.

\paragraph{Architecture:}
As shown in \figref{INTP}, the auto-decoder receives the latent code $\mbz$, the condition vector $\mbc$ and the current position $\mbx$ as the inputs and outputs the three field values $D_t(\mbx)$, $D_o(\mbx)$ and $D_{toa}(\mbx)$. The latent code $\mbz \in \mathbb{R}^{128}$ encodes the environment, the object geometry as well as the hand trajectory, and is computed by optimization both during training and testing. The condition vector $\mbc \in \mathbb{R}^4$ describes the hand state tokens from the real-time controller composed of the height of the goal object, the label of the currently controlled hand (left or right), the goal action label (grasping or placing), and the label of the current trajectory segment (approaching or leaving).
The composition of this information provides the semantic and spatial information that influences the trajectory. Together with the environment geometry, most indoor pick-and-place motion conditions can be described.

\paragraph{Training:}
We concatenate the latent code $\mbz$ with the condition $\mbc$ and the querying position $\mbx$, and optimize the network parameter $\theta$ and $\mathbf{z}_{i}$, which represents the $i$-th captured scene. 
At the training stage, the system
learns the mapping from the 
condition $\mbc$ and query position $\mbx$ to the $\{D_{t}, D_{o}, D_{toa}\}$ features: 
\begin{equation}
	\begin{aligned}
		\mathcal{L}_{\text{Rec}} = \sum_{\mbx \in E} \bigg( &\left\|f_{\theta}(\mbx, \mbz; \mbc)_0 - D_{t}(\mbx)\right\|_1 + \left\|f_{\theta}(\mbx, \mbz; \mbc)_1 - D_{o}(\mbx)\right\|_1 + \\
		&\left\|f_{\theta}(\mbx, \mbz; \mbc)_2 - D_{toa}(\mbx)\right\|^2_2 \bigg).
	\end{aligned}
\end{equation}
Note $L_{2}$ loss is used for $D_{toa}$ while $L_{1}$ loss is used for $D_{t}$ and $D_{o}$:
This is because $L_{2}$ loss is more sensitive to the nonuniform distribution caused by the anisotropic propagation of $D_{toa}$ and thus suitable.

We then train the network with all $N$ approaching/leaving scenes by following the maximum log posterior with respect to the latent code $\mbz$ and the network parameters $\theta$ as \cite{park2019deepsdf}:

\begin{equation}
	\underset{\theta,\left\{\mbz_i\right\}_{i=1}^N}{\arg \min } \sum_{i=1}^N\left( \mathcal{L}_{\text{Rec}}+\frac{1}{\sigma^2}\left\|\mbz_i\right\|_2^2\right)
\end{equation}
where $\sigma = 0.01$ as in DeepSDF~\cite{park2019deepsdf}. 

\paragraph{Runtime:} 
We estimate the hand motion by reconstructing the $D_{toa}$ field whose gradient corresponds to the hand trajectory; this is done by optimizing $\mbz$ such that the field reconstructed through $f_\theta$ matches the partially observable field values: 
\begin{equation}
	\begin{aligned}
		\hat{\mathbf{z}} = \underset{\mathbf{z}}{\mathrm{argmin}} \Bigg(& \sum_{\mbx \in E_{t}} \left( \|f_{\theta}(\mathbf{x,z};\mbc)_0 - D_{t}(\mbx)\|_1 + \|f_{\theta}(\mathbf{x,z};\mbc)_1 - D_{o}(\mbx)\|_1 \right) \\
		&+ \gamma \sum_{\mbx \in \mbx_{prior}} \|f_{\theta}(\mathbf{x,z};\mbc)_2 - D_{toa}(\mbx)\|^2_2 + \frac{1}{\sigma^2} \| \mbz \|^2_2\Bigg)
	\end{aligned}    
\end{equation}
where 
$\mbx_{prior} = \{\mbx_{O},\mbx_{T}\}$
represents the set of locations that $D_{toa}$ is known, including 
\begin{itemize}
	\item $\mbx_{O}$: $\mbx\  inside \  M_{O}$, such positions shall never be approached by the hand, thereby $D_{toa} = 0$;
	\item $\mbx_{T}$: the set of $\mbx$  when the grasping/placing is in action. Given $\mbc$ and $D_t$, a grasping/placing motion clip is matched in the database (described in~\secref{bimanual}), whose wrist trajectory is defined by $\tau_{prior}$.
    We include $\mbx\  \in \ \tau_{prior}$ where 
	$d(\mbx, M_{t}) < 5cm$ in $\mbx_{T}$,
\end{itemize}
$E_t$ is an object-centric grid aligned with the surrounding furniture, and 
$\gamma$ re-weights the loss of $D_{toa}(\mbx)$ to balance it with the other two losses according to the number of samples in the grid: $\gamma=N(E_{t})/N(\mbx_{prior})$.
The optimized latent variable $\hat{\mathbf{z}}$ and the decoder $f_\theta$ together form an implicit neural function that generates a wrist trajectory, taking into account its spatial relationship with the target object and surrounding obstacles.

During the goal approaching, the wrist moves along the descending gradient of the estimated time-of-arrival field $\hat{\phi} = 1/\hat{D}_{toa}$, and we can acquire the wrist velocity $\mathbf{v}$ at each $\mbx$ by:
\begin{equation}
	\mathbf{v}(\mbx) = -\nabla \hat{\phi}(\mbx)^{\circ -1},
\end{equation}
where ${}^{\circ -1}$ is the Hadmard inverse operation.
The velocity direction is determined by the inverse Hadamard product with the coordinate vector. Thereby, we can update the wrist position with $\mathbf{v}(\mbx)$. 
Because the goal grasping or placing position corresponds to a sink point of $\hat{\phi}$ where the gradient is zero, an integration starting from this point is unstable.
We thus find a position between the wrist position $\mbw$ in the initial frame receiving the instruction and the goal position contacting the object to start the trajectory integration:
\begin{equation}
	\mbx_{s} = \underset{\mbx} {\mathrm{argmin}} \{-\hat{\phi}(\mbx) + d(\mbx, \mbw) / \bar{v}\},\  s.t.\ \  \hat{\phi}(\mbx) \neq +\infty
\end{equation}
where $\bar{v}=1m/s$ is approximately the statistical average humanoid walking speed in a scalar, and the cost term $d(\mbx, \mbw) / \bar{v}$ encourages the hand to start/end the trajectory at a position closer to its current location. 
We then set $\mbx_s$ as the beginning/ending waypoint, and integrate from $\mbx_s$ by gradient descent to compute the positional elements of the newly planned tangent approaching/leaving trajectories of the wrist $\mbt'$, where the order of the leaving nodes is then reversed.

The 6D trajectory of the wrist $\tau_{plan}$ that follows the planned tangent trajectory $\mbt'$
is computed by utilizing the matched wrist motion $\tau_{prior}$, and then transferring 
the orientation $\mbR_{1:m}$ from $\tau_{prior}$ to that of the newly planned trajectory.
Specifically, the orientation of the $(m-i)$-th frame from $\tau_{prior}$ is added to the planning frame of $\mbx_{n-i}$, where $m,n$ are the frames synchronizing $\tau_{prior}$ and $\tau_{plan}$ based on the contact - the first contact frame for picking and release frame for placing.
Mathematically, the $3 \times 3$ rotation matrix $\mbR'_{1:m}$ of the $i$-th frame of the new trajectory $\tau_{prior}$ is computed by the following process: 
\begin{equation}
	\label{eq:6DTraj}
	\begin{aligned}
		\mbv & = \mbR_{m-i} (:,j) \times \mathbf{t}_{m-i}; \quad
		\theta_j = \arccos(\mbR_{m-i} (:,j) \cdot \mathbf{t}_{m-i}) \\
		\quad  \quad & \mbR^j_{align} = \mathbf{I} + [\mbv]_{\times} \sin(\theta_j) + [\mbv]_{\times}^2 (1 - \cos(\theta_j)) \\
		\quad \quad & \mbR'_{n-i}(:,j) = \mbR^j_{align} \mathbf{t'}_{n-i}  	 \quad \quad  (j=0,1,2)
	\end{aligned}
\end{equation}
where 
$[]_\times$ is the cross product operation matrix, 
$\mathbf{t}_{m-i}, \mathbf{t'}_{n-i}$ are the tangent vectors on $(m-i)$/$(n-i)$-th frames in $\tau_{prior}, \tau_{plan}$.
Furthermore, we linearly blend the final approaching frames on $\tau_{plan}$ towards $\tau_{prior}$ to ensure the object touching/releasing constraint is preserved. 

In summary, the matched prior determines the hand-object contact and realistic wrist orientations, and our runtime optimization identifies a latent code optimally tailored to the test environment (blue boxes in~\figref{INTP}), which is mapped to the time-of-arrival field that reliably extracts a collision-free positional path. Using Eq. \ref{eq:6DTraj}, a plausible 6D manipulation trajectory is produced.

\section{Deep Phase Interaction Controller}
\label{sec:InteractionController}
\begin{figure*}
	\includegraphics[width=\linewidth]{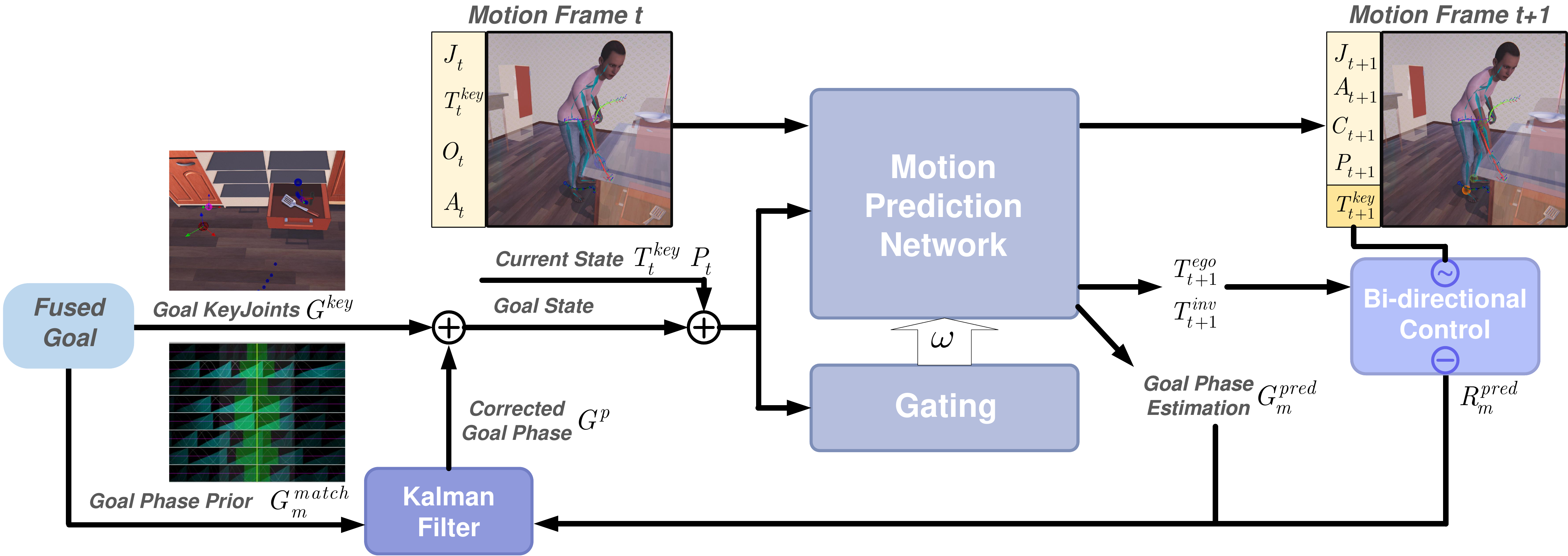}
	\caption{Framework of Our DeepPhase interaction controller:  The Kalman filter estimates the target phase correlated to the goal key-joint transformations. The Gating Network compares the features of key-joint transformations and the phase features from the current and goal frame, and after each motion prediction for the next frame, our bi-directional control blends the key-joint transformation prediction, and also feeds back the covariance to the Kalman filter based on the displacement of the bi-directional prediction.}
	\label{fig:MoE}
\end{figure*}

In this section, we describe our real-time motion generator that robustly tracks the sparse trajectories provided by the hand trajectory planner in \secref{INTP} and the root motion planned by a 2D planner (see~\secref{navigation}). This serves as the motion control stage in~\figref{pipeline} (dashed blue), provided with the goal states of the keyjoints, i.e., root, hip, and the two hands for guidance: these are planned by a high-level controller that is later described in \secref{highlevel}.

Our framework follows the DeepPhase controller~\cite{starke2022deepphase} but with extensions to control the hands and root of the body to move towards their goals, such as done in the Neural State Machine (NSM)~\cite{starke2019neural}.  
While the NSM requires all motions to be phase-labelled based on predefined phases, we instead make use of the phase labels computed by the Periodic Auto-encoder~\cite{starke2022deepphase} and develop an algorithm to guide the character towards the goal in the phase space (see \secref{goaldirectedcontrol}).    
Because the phase prediction at the goal is highly uncertain, we develop a scheme based on a Kalman filter to continuously improve the prediction over time (see \secref{kalman}). \figref{MoE} overviews how we combine these modules into a goal-driven control framework.

\subsection{Goal-directed Motion Prediction Network}
\label{sec:goaldirectedcontrol}
Our interaction controller employs a Mixture-of-Experts (MoE) architecture that auto-regressively predicts the state of the character for real-time control~\cite{zhang2018mode,starke2019neural,starke2022deepphase}; it receives the current pose of the character and the control signals, and outputs the pose in the next frame as well as the revised control signals
(see \cite{zhang2018mode,starke2019neural} for the details).  
We provide the goal states of certain keyjoints to guide the character towards the keyframe; in addition to the 2D root trajectories on the ground~\cite{starke2019neural}, the 6D transformations of the hip and left/right wrists are also given for precise pick-and-place motion synthesis. Further, the phase features at the goal states are given as additional control signals.
We list the details of the newly introduced input/output goal features below, and the remaining features can be found in our supplementary material.

{\it Input Features:} 

\noindent \textbf{Goal-focused Features} $\mbG_t^{keyJ} = \{\mathbf{T}_g^{hip}, \mathbf{T}_g^{root}, \mathbf{T}_g^{left}, \mathbf{T}_g^{right}\}$ set the goal-related transformation signals of the three key-joints together with the body root projected onto the ground, extracted from the goal time window pivoted on frame $g$, which is the next keyframe after current timestamp $t$. 
The keyframes are inserted when the following two types of events happen: (1) the action label of the body changes, such as when switching to ``walking'' from ``idling'', and (2) when the pick-and-place motion starts, ends and the contact state with the object changes in the middle.  


Each keyjoint goal $\mathbf{T}_g^{\{hip,root,left,right\}}$ contains the 2$s$ window ($w=13$) of the goal position and rotation centered at the next keyframe timestamp $g$.
This 2$s$ window may include more than one goal keyframe: for each current-pivoted timestamp $t_i \in (t-1, t+1)$, we define the corresponding goal timestamp $g(t_i)$ as the keyframe with the event next to $t_i$. Thereby, the goal time window may encompass multiple keyframes when $t+1 > g(t_i-1)$, informing the network about a transition towards the next goal.

\noindent \textbf{The Noised Phase Features at the Goal} $\mbP_g + \boldsymbol{\epsilon}$.  
The phase features at the goal keyframes are also provided as inputs, which help to align the timing of the training data~\cite{starke2019neural}, though here we use the phase by the DeepPhase~\cite{starke2022deepphase}.  
These are precomputed for the training data by the PAE and given as input.  
Gaussian noise $\boldsymbol{\epsilon} \in \mathcal{N}(0,0.1)$ is added to the phase features $\tilde{\mbP}_g = \{\tilde{\mathbf{p}}_g, \tilde{\mathbf{F}}_g\}$:
\begin{equation}
	\begin{aligned}
		\tilde{\mathbf{p}}_g & =
		\left( 
		\boldsymbol{A}_g \cdot \cos \left(2 \pi \cdot \boldsymbol{S}_g \right) + \boldsymbol{\epsilon}_r , \boldsymbol{A}_g \cdot \sin \left(2 \pi \cdot \boldsymbol{S}_g\right) + \boldsymbol{\epsilon}_i \right) \\ 
		\tilde{\mathbf{F}}_g & = \boldsymbol{F}_g + \boldsymbol{\epsilon}_f
	\end{aligned}
\end{equation} 
where $(\boldsymbol{A}_g, \boldsymbol{S}_g, \boldsymbol{F}_g)$ are the phase features at the goal composed of amplitude, phase, and frequency, and $\boldsymbol{\epsilon}_r, \boldsymbol{\epsilon}_i, \boldsymbol{\epsilon}_f$ are three independent Gaussian noise added on the real/imaginary parts of the phase and the frequency, respectively. The Gaussian noise added here will be equivalent to a Gaussian noise in the temporal domain after iFFT.

The purpose of adding the noise is to enhance the network robustness for precise goal tracking under various imprecise goal inputs, and simultaneously train the network to learn to rectify the noisy goal phase features to a predicted goal phase unbiased with respect to the goal keyjoint transformations. Actually, during testing, we cannot get prior knowledge of the phase features at the goal, while the network prediction learned from the data will help refine the goal phase features needed for tracking the keyjoint goals. We further develop an error-removal mechanism at the control scheme introduced in Section \ref{sec:kalman}.

{\it Output Features:} 

\noindent {\bf Egocentric Key Trajectory Prediction} $\mathbf{T}_{t+1}^{ego} = \{\mathbf{T}_{t+1}^{root}, \\  \mathbf{T}_{t+1}^{hip}, \mathbf{T}_{t+1}^{left}, \mathbf{T}_{t+1}^{right}\}$ are the future trajectories of the keyjoints in a 1$s$ future window.  

\noindent {\bf Goal-centric Key Trajectory Prediction} $\mathbf{T}_{t+1}^{inv} = \{\mathbf{T}_{t+1}^{root}, \\  \mathbf{T}_{t+1}^{hip}, \mathbf{T}_{t+1}^{left}, \mathbf{T}_{t+1}^{right}\}$ are the future trajectories of the keyjoints in 1$s$, but represented relative to the keyjoint's coordinate system at the goal: $\mbG_{t+1}^{keyJ}$. Ideally for each joint, the bidirectional prediction in the transformation matrix representation should follow the constraint: 
\begin{equation}
	\label{equ:BiDir}
	\mathbf{T}_{t+1}^{ego} = \mbG_{t+1}^{keyJ} \mathbf{T}_{t+1}^{inv}.
\end{equation}
Usually there is a discrepancy due to the lower precision of the prediction, especially when the goal is far away; therefore, we blend the two trajectories with increasing weight on the goal-centric prediction to encourage goal approaching.

\noindent \textbf{Final Goal Estimation} $\mbG_{t+1} = \{\hat{\mathbf{T}}_g^{hip}, \hat{\mathbf{T}}_g^{root}, \hat{\mathbf{T}}_g^{left}, \hat{\mathbf{T}}_g^{right}, \hat{\mbP}_{g}\}$: $\hat{\mathbf{T}}_g^{keyJ}$ are the network prediction of the future keyjoint transformations, in the form of a 2$s$ window pivoted at time $g$ with $w_g=13$ samples, and $\hat{\mbP}_{g}$ is prediction of the DeepPhase features at frame $g$.

    \begin{figure}[t]
		\includegraphics[width=\columnwidth]{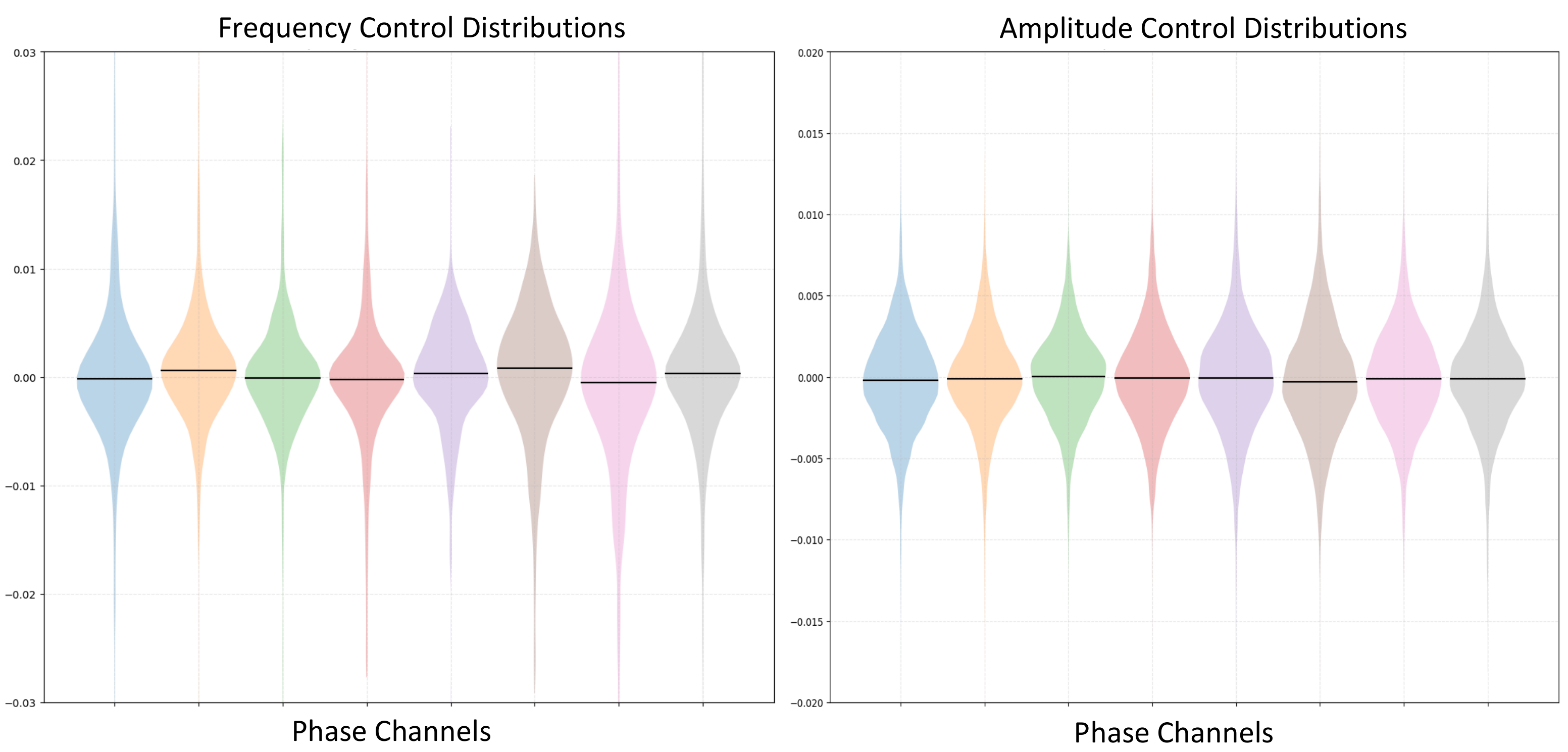}
		\caption{The distribution of amplitude and frequency control in $\mbU$, revealing the natural motion transitions, which exhibit consistent acceleration and deceleration patterns and perform the motion diversity during interactions, follows the zero-mean Gaussian distribution in the frequency-domain latent.}
		\label{fig:gaussianU}
    \end{figure}
    
\subsection{Goal-driven Control Scheme in Deep Phase Manifold}
\label{sec:kalman}

	In this section, we demonstrate that the DeepPhase control can be effectively represented as a linear dynamical system within the DeepPhase latent space. Leveraging this characteristic, we propose a goal-driven controller that enables smooth tracking of goal phase feature vectors in a latent space. The Periodic Auto-encoder of the DeepPhase framework converts high-dimensional nonlinear motion in the canonical space to a low-dimensional representation where linear dynamics can be applied, hence suiting learning long-term interactions. Additionally, we introduce a novel feedback loop by utilizing the network's predictive capability to estimate the goal phase, and establish a relationship between the measurement covariance provided to the Kalman filter and the bi-directional prediction bias in the spatial domain. This innovative approach significantly enhances guidance in the frequency domain latent, thereby improving the full-body motion coordination.

    \paragraph{DeepPhase as a Linear Dynamical System}
	Denoting the current phase states as $\mbX_t^p = [\mathbf{S_t}, \mathbf{A_t}, \mathbf{F_t}]^T$, the goal state as $\mbX_g^p$ and the body pose as $\mbY_t$, the DeepPhase~\cite{starke2022deepphase} system dynamics can be written as:
	\begin{equation}
		\begin{aligned} 
			& \mbX_{t+\Delta t}^p=\mbA \mbX_t^p+ \mbB \mbU \\ 
			& \mbY_{t+\Delta t}= f_{MoE} (\mbX_{t+\Delta t}^p, \mbY_t) \\ 
		\end{aligned} 
	\end{equation}
	where
	$\mbA, \mbB$ are constant square matrices 
	\begin{equation}
		\begin{aligned}
			\mbA & = \left[
			\begin{array}{ccc}
				1 & 0 & \Delta t \\ 
				0 & 1 & 0 \\ 
				0 & 0 & 1
			\end{array}
			\right], 
			\mbB = 
			\left[
			\begin{array}{ccc}
				\beta & 0 & 0 \\ 
				0 & 1 & 0 \\ 
				0 & 0 & 1
			\end{array}
			\right] 
		\end{aligned}
	\end{equation}
	and the network control $\mbU$ updates the phase variables as:
	\begin{equation}
		\label{eq:U}
		\begin{aligned}
			\mathbf{U} & = \left[
			\begin{array}{l}
				\Delta \mbS \\ 
				\Delta \mbA \\ 
				\Delta \mbF 
			\end{array}
			\right] =  f_{MoE}\left(\mbX_t^p, \mbX_g^p, \mbY_t\right)
            \sim \bm{C}[\mbX_g^p , \mbX_t^p]^T
		\end{aligned}
	\end{equation}
	where $\beta=0.5$ is a blending weight that enforces the phase update to follow the frequency. Among the state variables, the phase features are circularly updated and constrained within the range: the phase value $\mbS_t \in [0,1]$, the amplitude $\mbA_t \in [0,1]$, and the frequency $\mbF_t \in [0,\mbF_m]$, where $\mbF_m$ is the Nyquist frequency. Since $|\det(A)| = 1$ and the system is fully controllable via $\mbU$, this system exhibits critical stability, and with appropriate control inputs, the system state $\mbX_t^p$ can be guided to converge toward the desired goal state $\mbX_g^p$ in the latent space. 
    Because the controller network is trained to reduce the difference between the goal phase variables $\mbX_g^p$ and the current phase variables $\mbX_t^p$, we can approximate its output of phase update $\mathbf{U}$ by a linear transformation of the state error, i.e., $\mathbf{C}(\mbX_g^p-\mbX_t^p)$, with $\mathbf{C}$ as a coefficient matrix. 
    Due to the high-order smoothness of natural movements, the updates $\mathbf{U}$ are small and bounded.
    
    The control signal $\mbU$ can be statistically modeled as a zero-mean Gaussian distribution (see \figref{gaussianU}). 
    This assumption aligns with the characteristics of human motion sequences, which are often represented by Gaussian distributions~\cite{min2012motion,mukai2005geostatistical}.   
    As $U$ follows the Gaussian distribution, the process noise (error from directly inferring the next state by $\mbX_{t+\Delta t}=\mbA \mbX_t$) and the measurement noise from the controller network also follow the Gaussian distribution. The approximated linear dynamics of the phase state variables, coupled with the Gaussian noise assumption, allow us to use the Kalman filter to achieve stable goal tracking, as described next.

	\paragraph{Recursive Goal State Estimation by Kalman Filter}
	Ensuring precise and continuous target inputs $\mathbf{X}_g^p$ online is crucial for achieving robust goal tracking and precise convergence to the user's intended interactions. In our system, as presented in \secref{goaldirectedcontrol}, we assume that the goal phase is given. However, using the matched goal provides a biased phase prior due to differences in matching, and can only be provided for a short horizon around the action keyframe. To address this challenge and appropriately lead the character, we develop a framework for accurately estimating goal phases aligned with our planned hand trajectory using the Kalman filter. This filter is optimal for estimating the time-varying goal state under linear system dynamics with Gaussian noise. Our Kalman filter incorporates two measurements, namely predictions from the network ($m=pred$) which is unbiased and suits time-varying goals well, and the matched motion data prior ($m=match$) aligned with the same target action. Algorithm \ref{alg:kalman} outlines our pipeline.

	\begin{algorithm}[t]
		\DontPrintSemicolon
		\SetAlgoLined
		\KwIn{Motion frame $\bm{X}_t$, statistical variance $\bar{\bm{\sigma}}$}
		\KwOut{Goal phase states $\mbG^p$ and Covariance $\mbP$}
		
		Initialize $\mbG^p = 0$ and $\mbP = 0$\;
		\While{ $\mbG_t \neq$ null}{
			\If {MouseClicking}{
				Intialize goal $\mbG_{t} = \{\mbG_t^p, \mbG_t^{keyJ}\}$\;
			}
			Network Inference $\bm{X}_{t+1}, \mbT_{t+1}^{ego'},\mbT_{t+1}^{inv} = f_{MoE}\left(\bm{X}_t, \mbG_t \right)$\;
			Predict $\mbG_{t+1}^{p’} = F \mbG_t^p$ and $P_{t+1} = F P_{t} F^T + Q$\;
			Compute deviation scalar $d = Dev\left(\mbT^{ego'},\mbT^{inv}\right)$\;
			Define $R^{pred} = d Q = 0.2 d \bar{\bm{\sigma}}$\;
			$\{\mbG_m, \mbR\} = \{\mbG_m^{pred}, R^{pred}\}$ \;
			$hand$ := (is left-hand task ongoing) ? $left hand$ : $right hand$\;
			\If {$Nearby(X_t^{root}, G_t^{root})$ and $\neg Nearby(X_t^{hand}, G_t^{hand})$}{
				$\{\mbG_m, \mbR\} := \{\mbG_m, \mbR\} \cup \{\mbG_m^{match}, R^{match} = Q\}$\;
			}
			\For{each measurement $\{\mbG_m(i), \mbR(i)\}$}{
				Compute Kalman gain $\mbK = \mbP' \left(\mbP' + \mbR(i)\right) ^ {-1}$\;
				Update phase state $\mbG^p \leftarrow \mbG^p + \mbK\left(\mbG_m(i) - \mbG^{p'}\right)$\;
				Update covariance matrix $\mbP \leftarrow (\mathbf{I} - \mbK)\mbP$\;
			} 
			Update goals $\mbG_{t+1} = \{\mbG^p, \mbG_{t+1}^{key}\}$\;
			$t:= t+1$\;
		}
		\caption{Goal Phase Tracking with Kalman Filter}
		\label{alg:kalman}
	\end{algorithm}

	We first initialize the goal phase ${\mbG^{p}}$ 
	and its covariance $\mbP$
	using those estimated by the Kalman filter in the previous frame:
	\begin{equation}
		\begin{aligned}
			& \ \ {\mbG^{p}} \leftarrow \mbF \mbG^p, & \mbP \leftarrow \mbF \mbP \mbF^T + \mbQ
		\end{aligned}
	\end{equation}
	where
	\begin{equation}
		\begin{aligned}
			\mbF &= \begin{bmatrix} 1 & 0 & \Delta t \\ 0 & 1 & 0 \\ 0 & 0 & 1 \end{bmatrix}, & 
			\mbQ &= \begin{bmatrix} \sigma_s^2 & 0 & 0 \\ 0 & \sigma_f^2 & 0 \\ 0 & 0 & \sigma_a^2 \end{bmatrix},
		\end{aligned}
	\end{equation}
	$\sigma_s^2, \sigma_f^2, \sigma_a^2$ are the process noise of the phase value, frequency, and amplitude, each of which is normalized to 0.2$\times$ of their statistical variance of the entire dataset. 
	
	Then, the goal phase ${\mbG^{p}}$ and covariance $\mbP$ are refined iteratively with
	the individual measurements $m = \{pred, match\}$ by its Kalman gain $\mbK_m$ computed using the goal phase $\mbG^p_m$ and its covariance $\mbR_m$:
	\begin{equation}
		\begin{aligned}
			\mbG^p &\leftarrow \mbG^{p} + \mbK_m\left(\mbG_m^p - \mbG^{p}\right)  \\
			\mbP & \leftarrow (\mathbf{I} - \mbK_m)\mbP
		\end{aligned}
	\end{equation}
	where
	$		\mbK_m  = \mbP \left(\mbP + \mbR_m\right) ^ {-1} $.


	Regarding the measurement from the network, 
	the goal phase $\mbG_{pred}^p$ is simply that estimated by the MoE 
	and the covariance matrix is computed by scaling $\mbQ$ 
	by a confidence value $c$ that is based on the bidirectional prediction constraint in Eq. \ref{equ:BiDir}: the more the goal-centric prediction deviates from the ego-centric prediction, the less accurate the prediction can be. Thus we compute 
	$\mbR_{pred}$ by: 
	\begin{equation}
		\mbR_{pred} = c \mbQ = 0.2 c \bar{\bm{\sigma}},
	\end{equation}
	where $c$ is computed by summing the rotational cost in Frobenius norm and the positional cost for the three key joints, as:
	\begin{equation}
    \label{eq:dev}
    \begin{aligned}
        c &= Dev\left(\mbT_{t+1}^{inv}, \mbT_{t+1}^{ego}\right) \\ 
        & = \sum_{j = 0}^{3} \left( \alpha \| \log(\bm{r}_j^{ego^T} \bm{r}_j^{inv}) \|_F + \beta \| \bm{t}_j^{ego} - \bm{t}_j^{inv} \| \right).
    \end{aligned}
\end{equation}
where $Dev(*,*)$ computes the deviation of two 6D configurations and $\mbT_{j}^{ego} = \left[\bm{r}_j^{ego}, \bm{t}_j^{ego}\right]$, $\mbT_{j}^{inv} = \left[\bm{r}_j^{inv}, \bm{t}_j^{inv}\right]$ are the ego/goal-centric transformation prediction of the $j$-th key joint; the normalization parameters are set to $\alpha = 1/\pi$ and $\beta$ the reciprocal of the body width.
	Regarding the measurement from the matched motion in the database, the corresponding phase of each frame can be precomputed by the PAE, and the covariance matrix is set to $\mbQ$.
	
	Our Kalman filter optimally corrects the matched phase prior and simultaneously estimates a continuous-time goal trajectory in the latent space that infills the interaction keyframes, thereby enhancing the goal-driven motion quality as evaluated in \secref{ablationExp}.

	\section{Navigation and Bimanual Goal Scheduling}
	\label{sec:highlevel}

\begin{figure}[t]
		\includegraphics[width=\columnwidth]{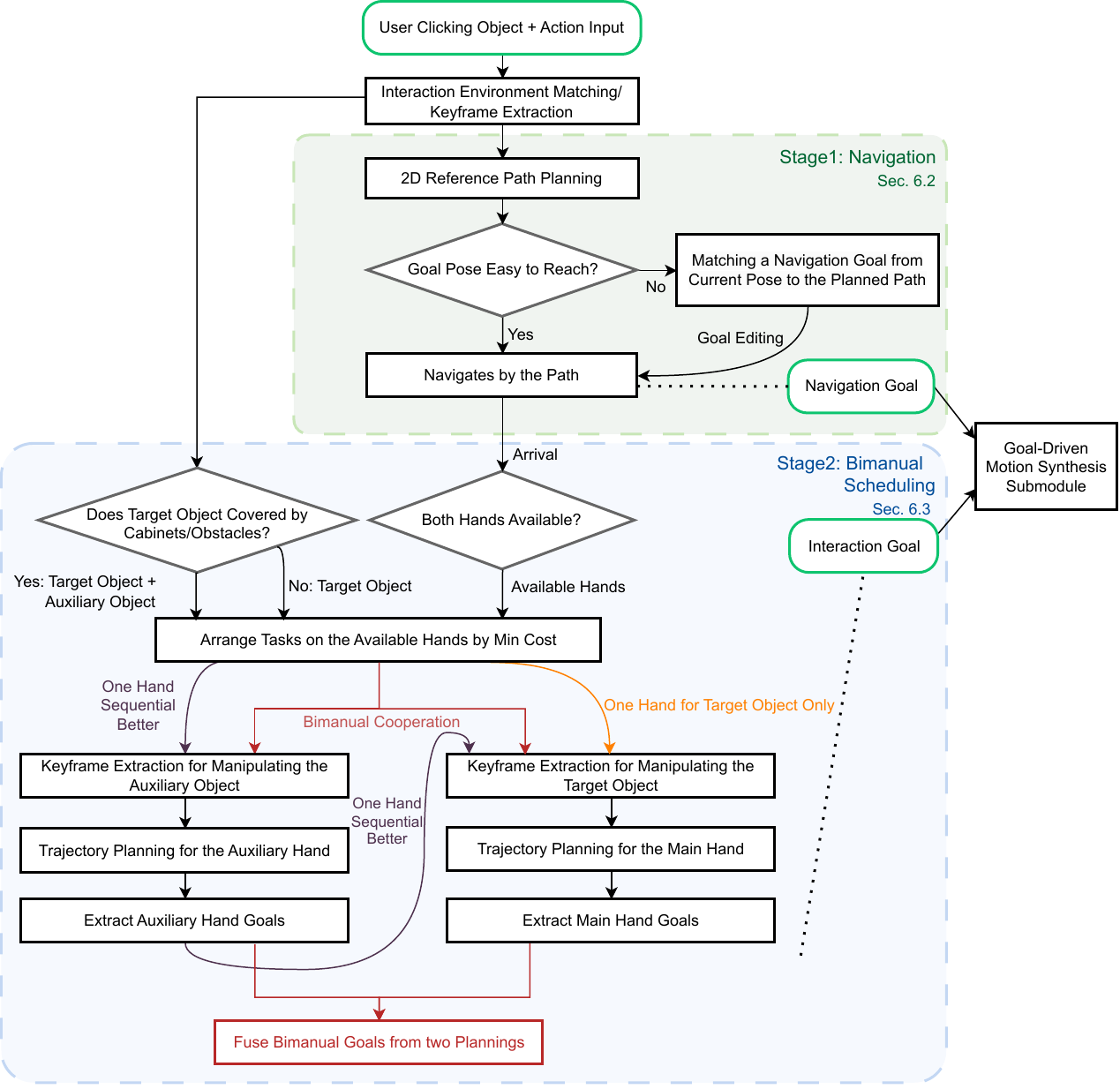}
		\caption{Overview of our state-machine structure for synthesis coordinate interaction-guiding goals (corresponds to~\figref{pipeline}, the goal coordination block before getting the fused goal). It adaptively allocates coordinated goals for both the hands and the entire body according to the test environment as described in~\secref{bimanual}. The navigation goal before arrival (see~\secref{navigation}) and the interaction goals during manipulation (plot in green capsules) are sequentially generated to guide the DeepPhase interaction controller.}
		\label{fig:stateMachine}
	\end{figure}

We now describe how we navigate the character to the target object and then schedule the bimanual interaction goals for pick-and-placing. This procedure can be decomposed to a goal matching subsystem (see~\figref{pipeline} green sub-system) and a state machine (see~\figref{pipeline} grey block of Goal Coordination). Before running the goal matching, we first compute a feature of the environment based on 3D CNN (see \secref{goalMatching}).
Next, we use this feature to retrieve and stitch a series of locomotion snippets, driving the navigation to the goal object (see 
~\secref{navigation} and~\figref{pipeline} cyan block). Once the character arrives, the bimanual goal matching module (see \secref{bimanual} and~\figref{pipeline}, green block) plans a sequence of keyframes to execute the pick-and-place actions. \figref{stateMachine} overviews the logic of our goal coordination module in the form of a state-machine structure that produces coordinated full-body motion.

	\begin{figure}[t]
		\includegraphics[width=.7\columnwidth]{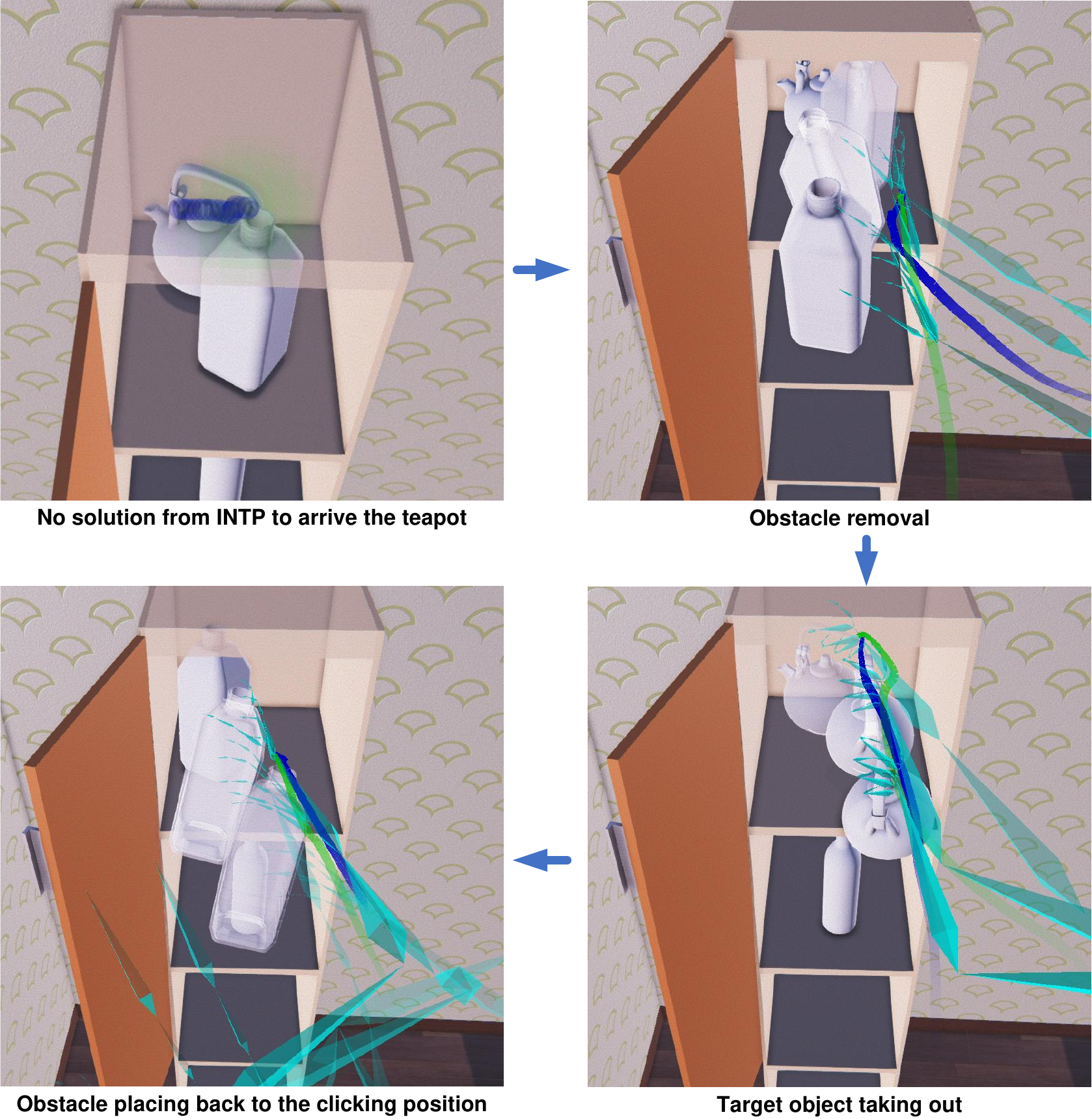}
		\caption{Our trajectory planner will detect the cases where the clicking target (teapot) cannot be grasped. To always achieve the interaction task, our system can sequentially remove the obstacle and then re-plan to take the target object out utilizing the available hands, and finally, we can place the obstacle back to any clicking position inside the blocks. Among the trajectory visualizations, the object/skeleton from transparent to opaque sequentially shows the in-hand motion; the blue and green trajectories record the approaching and leaving motion, respectively.}
		\label{fig:failPlan}
	\end{figure}

	\subsection{Environment Features}
	\label{sec:goalMatching}
	The goal matching mechanism is to provide a motion prior for grasping/placing an object; to achieve a quick matching, we compute a volumetric scene feature around the goal position by 3D CNN. An auto-encoder structure with 3 layers of convolution + max-pooling, followed by 3 layers of deconvolution + upscaling, is designed. 
	A distance field of the environment, pivoted at the object center and computed by fast marching, is fed into the network as an input.  
	We extracted the flattened latent code $\bm{z} \in \mathbb{R}^{16}$ together with the hand manipulation state $\bm{c}$ described in Section \ref{sec:INTP} as the matching feature vector from all the goal keyframes in the training set and store them in a KD tree structure.
	
	During testing, we encode the environment around the target object $o$  to form its matching feature with the manipulation state $c$:
    $f = \{z_{o}; \bm{c}\}$ and match $K-$nearest neighbors in the KD tree.
	
	\subsection{Navigation Matching}
	\label{sec:navigation}
	We now describe our scheme to guide the character in a current pose to arbitrary goal locations in the scene.  Solving such a problem only by motion matching is not feasible, especially if the goal location is far away.  We cope with this issue by first planning a motion by 2D navigation with breadth-first search, and then finding a sequence of matched motions that can be stitched to reach the goal.

	\paragraph{Computing the 2D Path} 
	The 2D path from the current position of the character $\bm{p'}$ to the goal position and direction $(\bm{p}_g^{root}, \bm{d}_g^{root})$, which can be acquired from the matched destination keyframe with hand-object operation, is computed by a breadth-first search. The cost function for each position $p$ is defined as follows:
	\begin{equation}
		c(\bm{p}) = \begin{cases}
			\infty, & \textit{if} \  \left(\bm{p}\right) in \ obstacles \\
			angle(\bm{p}-\bm{p}_g^{root}, \bm{d}_g^{root}) / \pi + 0.5, & \textit{if} \  ||\bm{p}-\bm{p}_g^{root}||_1 < l \\
			1,  & \ \textit{otherwise}
		\end{cases} 
	\end{equation}
	where $l$ is a threshold set to $0.4m$; the second cost encourages the character to reach the goal from a direction that is parallel to the facing direction of the character at the goal.  The optimal path is computed using the fast marching method. 
	
	
	\paragraph{Following the 2D Path by Matching}
	Given a 2D path computed by breadth-first search and the current pose of the character $\bm{p'}$, goal waypoints of the three keyjoints are matched from the dataset to synthesize natural locomotion that follows the path.  As it will be difficult to find a match for a very long path, we track the path in segments by querying a series of goal positions uniformly sampled within each $3m$ segment from the start root position. In each segment, the keyjoint goal with the least deviation and without any collision with the surrounding obstacles is then selected.
	
	For preparing a motion dataset,  navigation features defined by  
	$\bm{N} = \{\bm{d}_g^{root}, \bm{p}_g^{root}, \bm{r}^{left},\bm{p}^{left}, \bm{r}^{right},\bm{p}^{right}, \bm{r}^{hip}, \bm{p}^{hip}, \bm{v}^{root}\}$ are computed for every motion between adjacent keyframes in the database, where $\bm{d}_g^{root} \in \mathbb{R}^{2} $ and $\bm{p}_g^{root} \in \mathbb{R}^{2}$ are 2D orientation/position at the goal and $\bm{r}^{keyJ} \in \mathbb{R}^6$, $\bm{p}^{keyJ} \in \mathbb{R}^3$ ,
	$\bm{v}^{root} \in \mathbb{R}^2$
	are keyjoint orientation/position and root velocity relative to the root coordinate at the beginning of each motion clip.   These features are used as the key to search for motion clips during navigation.

	\subsection{Bimanual Task Scheduler}
	\label{sec:bimanual}
	
	Humans use both hands to achieve pick-and-place tasks efficiently in living environments, e.g., using one hand to open the door of the cabinet while taking out an object with the other hand; to produce such movements, we propose a bimanual task scheduler that schedules the optimal series of actions by the full body.
	
	The pipeline of the bimanual task scheduler proceeds as follows:
	
	\noindent
	{\bf 1. Detecting the target and auxiliary objects: } 
	The user clicks the target object with the mouse, and then the system detects the target object (denoted as $o_g$) and all the other objects involved, which we call auxiliary objects $\{o_{d}\}$, such as the obstacles, and the container that contains the target object (whose manipulable part like a door/drawer is also denoted as $o_{d}$) by ray-casting.

    The system then checks if any auxiliary object needs to be manipulated. In addition to handling the closed cabinets, Fig.~\ref{fig:failPlan} shows another no-solution case to directly take out the inner teapot, where our system automatically controls the character to remove the obstacle object before taking the target one. This is determined by whether the target object can be taken outside the cabinet without collision, based on the planned trajectory.
	
	\noindent
	{\bf 2. Matching the Motion for Main/Auxiliary Tasks: }
	Next, the hands to grasp the auxiliary object and the target object are scheduled, and the poses of the three keyjoints, including the left hand, right hand, and the hip, are extracted from the dataset where the character grasps the auxiliary object $o_{d}$, and the target object $o_g$.   
	
	The task scheduling is done according to the availability of the hands: 
	If both hands are free, the goal prior matching algorithm described in Section \ref{sec:goalMatching} is conducted separately for $o_g$ and $o_{d}$, resulting in 4 pairs of hand-object interaction matches denoted by $\{M(H_{\{l,r\}}, o_g)$,
    $M(H_{\{l,r\}}, o_{d})\}$ where $\{l,r\}$ represents the left and right hand. If only one hand is free, the opening and grasping tasks are allocated sequentially to the free hand.
    
    If two hands are free, the total goal costs are then computed for all potential schedules, including sequentially opening and grasping with either hand and co-temporally opening and grasping with two hands. 
	The keyframes to grasp the auxiliary object and the target object are matched such that they minimize the designed deviation cost $C$ for all possible combinations of keyframes for grasping the auxiliary object and the target object:  
	\begin{equation}
		\label{eq:cost}
		\begin{aligned}
			\min C = \min_{i_d,i_g} \sum_{k=1}^3 \{ Dev(M_k(H_{i_d}, o_{d}), \bm{p}^{key}_k)  \\
			+ w Dev(M_k(H_{i_d}, o_{d}), M_{k}(H_{i_g}, o_{g}))
			\}
		\end{aligned}
	\end{equation}
	where $M_k$ is the matching function that returns the 6D transformation for each key joint $k$ (left hand, right hand, and root) for the matched keyframe, $i_d,i_g$ are the hand indices used for grasping the auxiliary and target objects, 
	$\bm{p}^{key}_k$ is transformation of key joint $k$ in the current frame, and $w$ is a weighting set to 2.

	\noindent
	{\bf 3. Runtime Motion Goal Synthesis: }
        \label{sec:MotionGoalSynthesis}
	Given a sequence of grasping/releasing keyframes extracted from the dataset, we adjust their corresponding keyjoint transformations to the runtime scene after a collision-solving process that pushes back any penetrations of the hip or the idle hand along the penetration normal. The adjusted goal on the hand-object contact keyframe is then passed to the trajectory planner to generate the hand trajectories, which we compose with the other goal keyjoints to drive the DeepPhase interaction controller. We finally establish a state machine for these procedures, including the navigation stage before scheduling the manipulation, as shown in~\figref{stateMachine}.

	\section{Data Capture and Labeling}
    \label{sec:dataLabeling}
    \begin{figure*}[t]
		\includegraphics[width=\linewidth]{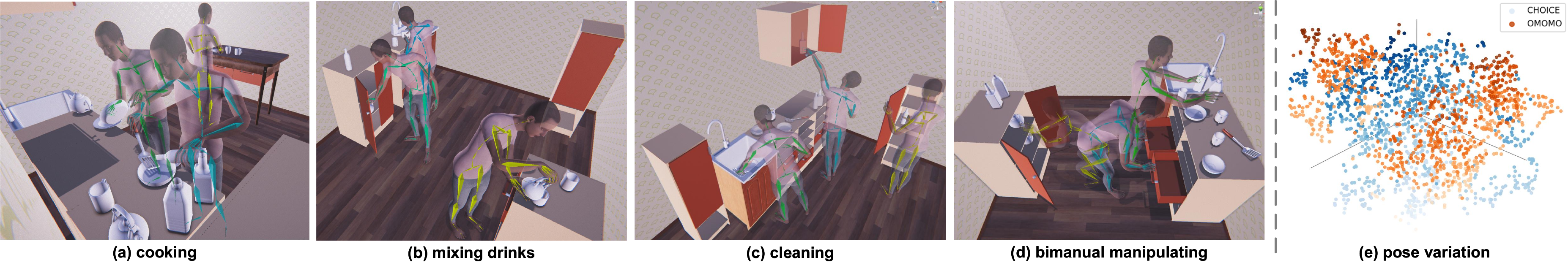}
		\caption{Snapshots of our CHOICE dataset, which incorporates different tasks like (a) cooking with ingredients and seasoning, (b) making milk tea and pouring into multiple mugs, (c) cleaning the table, and (d) other bimanual manipulations that co-temporally operate two objects. In (e), we visualize the pose variations in datasets after T-SNE clustering.}
		\label{fig:visData}
	\end{figure*}

	
	\begin{table}
		\caption{We list the distribution of action labels and task descriptions for the key joints of our dataset, and evaluate the motion diversity and smoothness by class. "Cook" denotes the motion of operating the pan/spatula, and "Pour" encompasses both pouring drinking and the liquid ingredients. The "Div." metric evaluates the motion diversity by Average Pairwise Distance (APD) in $cm$. We sample and compute this metric consistently with HuMoR~\cite{rempe2021humor}, where its motion latent reconstructs a 95 APD from the huge AMASS dataset~\cite{AMASS:ICCV:2019}. We also measure the motion smoothness by reporting the mean acceleration (in $cm/s^2$) and jerk (in $m/s^3$) averaged among all non-finger joints.}
		\label{tab:dataDistribution}
		\begin{tabular}{l|l|r|r|r|r|r}
			\hline
			Action     & Task             & Sec. & \% & Div. & Acc. & Jerk\\ \hline
			& Grasp/Place      &   3451      &  12.7 & 96.2 & 90.7 & 62.1 \\
			Object & Cook             &  2845       &  10.5 & 91.1 & 76.4 & 57.1 \\
			& Drink            &    98     &  0.4 & 88.1 & 89.1 & 50.7 \\
			& Pour             &    3075     &  11.3 & 74.8 & 86.1 & 49.9 \\ \cline{1-2}
			Door      & Open/Close &    3046     & 11.2 & 100.3  & 87.4 & 50.9 \\
			Draw       & Drag/Push      &     1731    & 6.4 & 130.5  & 80.6 & 50.5\\ \hline
			Idle       & Stand/Squat &     23352    & 85.9 & 95.8 & 92.9 & 55.5\\
			Walk       & Approach/Leave         &     3845    &  14.1 & 84.2 & 146.2 & 62.4 \\ \hline
		\end{tabular}
	\end{table}

	\begin{table}
		\caption{Dataset evaluation on grasping precision. Our evaluation is computed by averaging the maximum penetration volume and depth, and floating displacement during each pick-and-place.}
		\label{tab:dataPrecision}
		\begin{tabular}{c|c|c}
			\hline
		p.vol. $(cm^3)$            & p.dist. $(cm)$ & disp. $(cm)$\\ \hline
		     1.74   &   0.22    & 0.61  \\
			\hline
		\end{tabular}
	\end{table}

	We extensively captured and post-processed 150 motion sequences using the Vicon Sh\'{o}gun~\footnote{https://www.vicon.com/software/shogun/} motion capture system. 
	The dataset comprises long-term, delicate interactions in lifelike scenes, such as a corridor with a drinking bar (\figref{visData}(d)), and a well-equipped kitchen room (\figref{visData}(a)). We approximate various involved furniture by configuring up iron frames, the surfaces in drawers and tables are formed by wired mesh, and all the articulated doors are made of glass. This setup ensures that every contained object is visible to the surrounding cameras. After filtering the initial data, we post-process our captures in Vicon Sh\'{o}gun by minimizing finger-object penetration and ensuring contact during grasping. We fit an SMPL~\cite{loper2023smpl} body model to the subject using SOMA~\cite{SOMA:ICCV:2021}, and export it to the Vicon Sh\'{o}gun body template with a hand skeleton.
    
    The breakdown of our dataset is shown in~\tabref{dataDistribution}:
    Following the detailed MoCap procedure described in our supplementary, our dataset achieves 
     high smoothness, as detailed in~\tabref{dataDistribution}, along with precise manipulation, as shown in~\tabref{dataPrecision}.

    Each motion sequence in our captured dataset achieves a high-level semantic goal, e.g., cooking a seasoned steak, making milk tea and drinking, or cleaning a crowded table, with an average duration of 3 minutes.
    The dataset exhibits substantial diversity in terms of hand movements, object types (including 14 pieces of heavy furniture, 12 freely movable objects, 6 doors, and 4 drawers), and spatial layouts, which vary across cabinets and drawers at different widths and heights, object-object relations, obstacles, and placement heights ranging from 20$cm$ to 200$cm$. Our dataset offers pose diversity comparable to the OMOMO~\cite{li2023object} dataset (see~\figref{visData}(e)), and captures rich motion variations for each action category, as quantified in the "Div" column of ~\tabref{dataDistribution}.

	In addition to the sequence-level labelling, we annotate each frame of the three key joints with action labels. The hip is labelled as either "idle" or "walk", and each hand is labelled with actions like "open a door", "open a drawer", or "approach an object", along with a contact label with the object. A combination of labels for the three keyjoints effectively describes the current action type (see~\tabref{dataDistribution}).

    We further utilize the PAE's encoder to extract the phase latent vector that implicitly represents the full-body motion information of the $[-1s,1s]$ time window pivoted on each frame. Consequently, each frame is described by its current state and the next goal state, with each state consisting of its phase latent vector, action labels, and the three key-joint transformations.
	
	\section{Experiment \& Evaluation}
	We first describe the experimental setup for model training/testing, and then qualitatively and quantitatively evaluate our methods.

	\subsection{Experiment Setup}
	\paragraph{Layout Data Preparation and Trajectory Planner Training}
	We implement our interactive character control system in Unity 3D engine and update the motion with 60 fps on a computer equipped with 5.1GHz Intel i5-13600KF and NVidia GeForce RTX 4090. For training the neural implicit trajectory planner, we reduce the network size from~\cite{park2019deepsdf} to a 6-layer MLP with 256 neurons per layer and a skip connection to the 4th layer. We prepare 5546 interaction cases, each of which is composed of a pick-and-place motion and the local environment geometry represented by position-field feature pairs (~\secref{INTP}) uniformly sampled in a resolution of 2.5cm within a $(80\text{cm})^3$ cubic region around the target object.
    
    Training the auto-decoder using 90\% of our data takes 8 hours on two Tesla A100s for 300 epochs. At runtime, each trajectory planning task optimizes the latent code by 100 steps on our desktop, taking about 1.25 seconds.
	
	\paragraph{Interaction Controller Network Training}
	For motion data processing, we extract 8 channels in the frequency domain with the same network structure as~\cite{starke2022deepphase} to train the Phase Auto-encoder. After training it with 120 epochs for 10 hours on the training split, we pre-process the full motion dataset with DeepPhase features. To achieve a real-time inference over 60 fps, we design the DeepPhase interaction control network with 3 layers of 128 neurons in the gating network and the main motion prediction network with 3 layers of 512 neurons. We train the controller network using 70\% of the dataset with 150 epochs, which takes approximately 8 hours.
	
	\subsection{Trajectory Planner Evaluation}
	
	We compare our implicit neural trajectory planner with current state-of-the-art planners, including both data-driven methods and optimization-based robotic planners. We analyze the trajectory quality and evaluate the generalization capability of the planners to unseen environments.

	\begin{figure}[t]
		\includegraphics[width=.9\columnwidth]{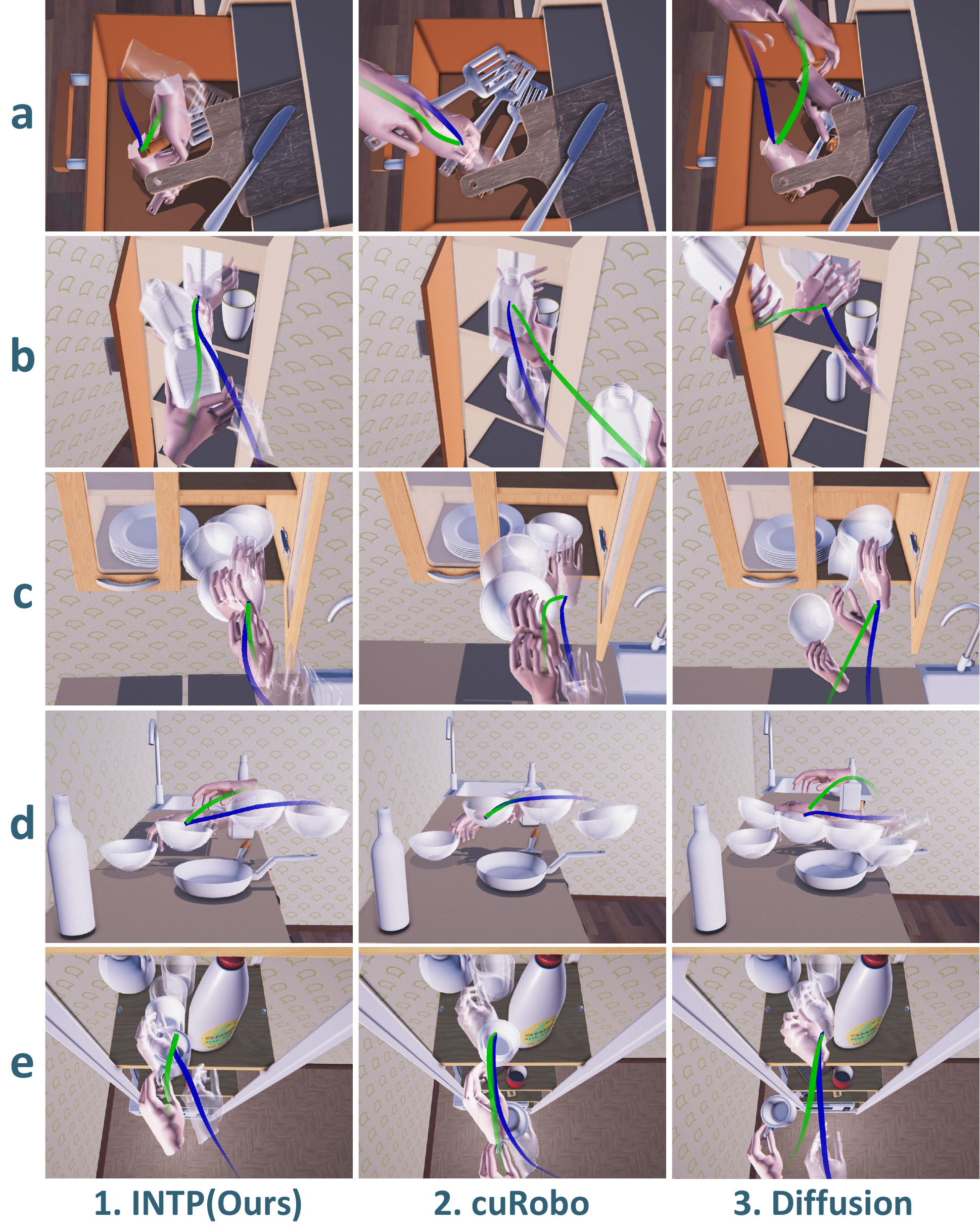}
		\caption{Comparison of trajectory planners under representative generalization scenes with narrowed cabinets and expanded obstacles. The blue and green curves shown in two views record the approaching and leaving trajectory of the wrist, respectively. See more comparisons in our supplementary video.}
		\label{fig:visCompareGene}
	\end{figure}
	\subsubsection{Comparison on Motion Realism}
	
	We compare our trajectory planner with two state-of-the-art trajectory planners (see \figref{visCompareGene}): the parallelized collision-free minimum-jerk model proposed in cuRobo~\cite{sundaralingam2023curobo}, 
    for robotics applications; and the Diffusion Policy~\cite{chi2023diffusion}, a generative data-driven planner trained with the same set of motion data by 6000 epochs, taking $~\sim$2 days on 2 Tesla V100 GPUs.

    \paragraph{Comparison to cuRobo} The cuRobo planner aims to produce collision-free motion with an optimal jerk cost; however, its wrist trajectory deviates from realistic human motion in two aspects.
    
   First, cuRobo tends to produce straight trajectories, where the hand and the in-hand object move too close to obstacles (see \figref{visCompareGene}(b,c-2)). 
   Also, as cuRobo is designed for rigid-body grippers, the object tends to produce minimal rotation after grasping, again resulting in little space between the object and the environment.
   In contrast, our planner, learned from kinematic features, employs dexterous strategies to produce human-like trajectories;
   for instance, when grasping a spatula out from a drawer, it makes use of the finger degrees of freedom for rotating the in-hand object while pulling it out vertically, thereby reducing the swept volume (see \figref{visCompareGene} (a-1)).

   Secondly, cuRobo can only produce hand trajectories with a fixed base, resulting in a lack of coordination of the locomotion with the pick-and-place task; including the base translation and orientation with the minimum jerk criteria would be difficult, as human models require stepping, causing many local minima. Indeed, producing human-like motion with optimization in canonical space requires a significant amount of cost/reward tuning in high dimensions, which is not practical.
   In contrast, our data-driven model motion can produce well-coordinated motions that are well-suited for the condition represented by the latent code  $\mbz$ and the condition parameters $\mbc$, eliminating the fixed hand base to achieve body-level guidance during its trajectory prediction.

    \paragraph{Comparison to Diffusion Policy}
    Although learned from the same kinematic dataset, we find that the Diffusion Policy struggles to robustly produce realistic trajectories in test scenes due to the sparse dataset with high diversity. 
    It suffers from over-straight trajectories (see~\figref{visCompareGene}(a,b-3) blue curves), irregular wrist orientation (see~\figref{visCompareGene}(b,c-3) and sudden change in height that result in heavy placement
    (see ~\figref{visCompareGene}(d-3)).

	\begin{table}[]
		\caption{Ablation study and comparison to the human motion data. The unsmoothness is computed by averaging the accelerations among all joints in $cm/s^2$. The sliding sums up the two foot velocities during ground contact by $cm/s$. The rooted mean-square curvature (RMSC) of the wrist trajectory is averaged among all the approaching/leaving cases in $cm^{-1}$.} The Fr\'{e}chet distance (FD) evaluates the motion similarity to the dataset by extracting each 2s sliding time-window poses as the feature vector. And the FD and unsmoothness are evaluated after aligning the joint transformations in their root-centric local coordinate systems.
		\label{tab:ablation}
		\begin{tabular}{l|r|r|r | r}
			\hline
			& Unsmth. $\downarrow$ & Slid. $\downarrow$ & FD $\downarrow$ & EE RMSC$\downarrow$ \\ \hline
			w/ cuRobo &      6.54     &    7.22   &  157 & 0.0725\\ \hline   
			w/o any goal phase          &    6.78       &    7.55 & 166 & 0.0982\\      
			w/o predicted goal      &         6.36          &  6.80  &  145 & 0.0721 \\
			w/o matched goal &         6.30          &  6.86  & 160 & 0.0614\\
			ours   &        \textbf{5.94}          & \textbf{6.17}   & \textbf{139} & \textbf{0.0509} \\ \hline  
			
			Dataset &      2.22   &  1.02      &  -  & 0.0408 \\ \hline
		\end{tabular}
	\end{table}
    
    {\it Quantitative Evaluation:}
    The quantitative results in~\tabref{ablation} also validate that our implicit neural trajectory planner can produce body motion more similar to human movement, resulting in an overall 11\% lower Fr\'{e}chet distance to the dataset compared with the motion following the cuRobo trajectory. Moreover, the improved motion smoothness and reduced foot sliding illustrate that our generated trajectory can be better tracked by the character, enabling it to achieve interactions with higher body coordination, benefiting from the aforementioned strengths of our kinematic data-based planner.

	\begin{figure}[t]
		\includegraphics[width=\columnwidth]{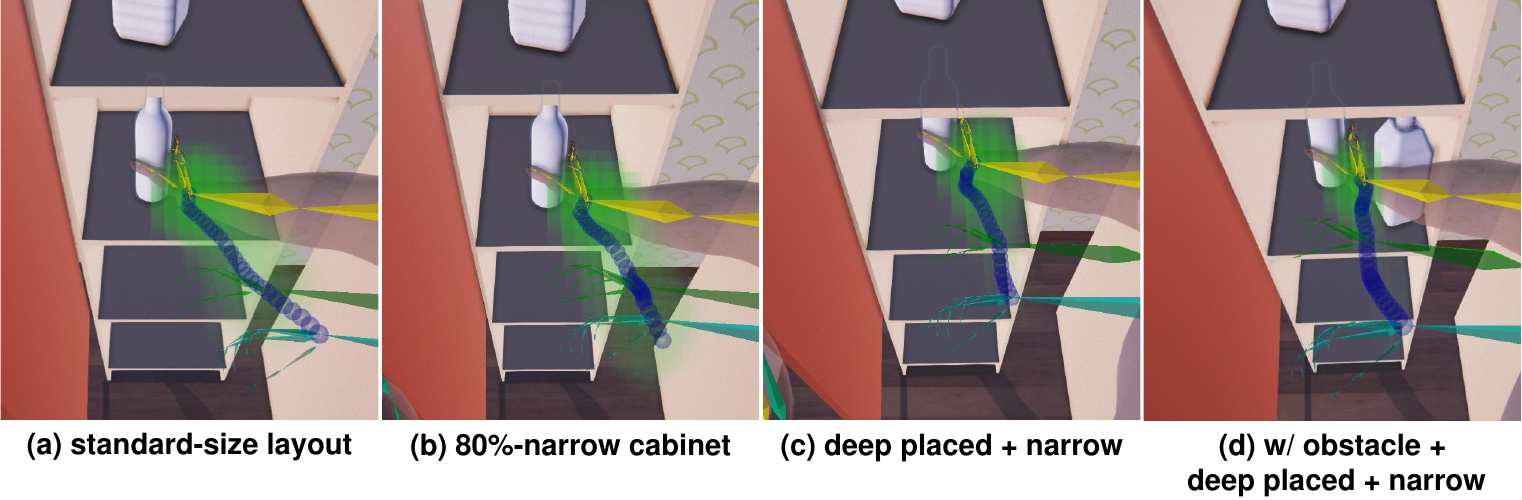}
		\caption{Figures from (a) to (d) sequentially show the planning of approaching a bottle with field deformation under a standard-sized shelf, an 80\% narrow (29.5$cm$ in width) shelf, a deeply placed case in the narrow shelf, and a deeply placed case with another bottle in a narrow shelf.}
		\label{fig:visGene}
	\end{figure}

	\begin{figure}[t]
		\includegraphics[width=\columnwidth]{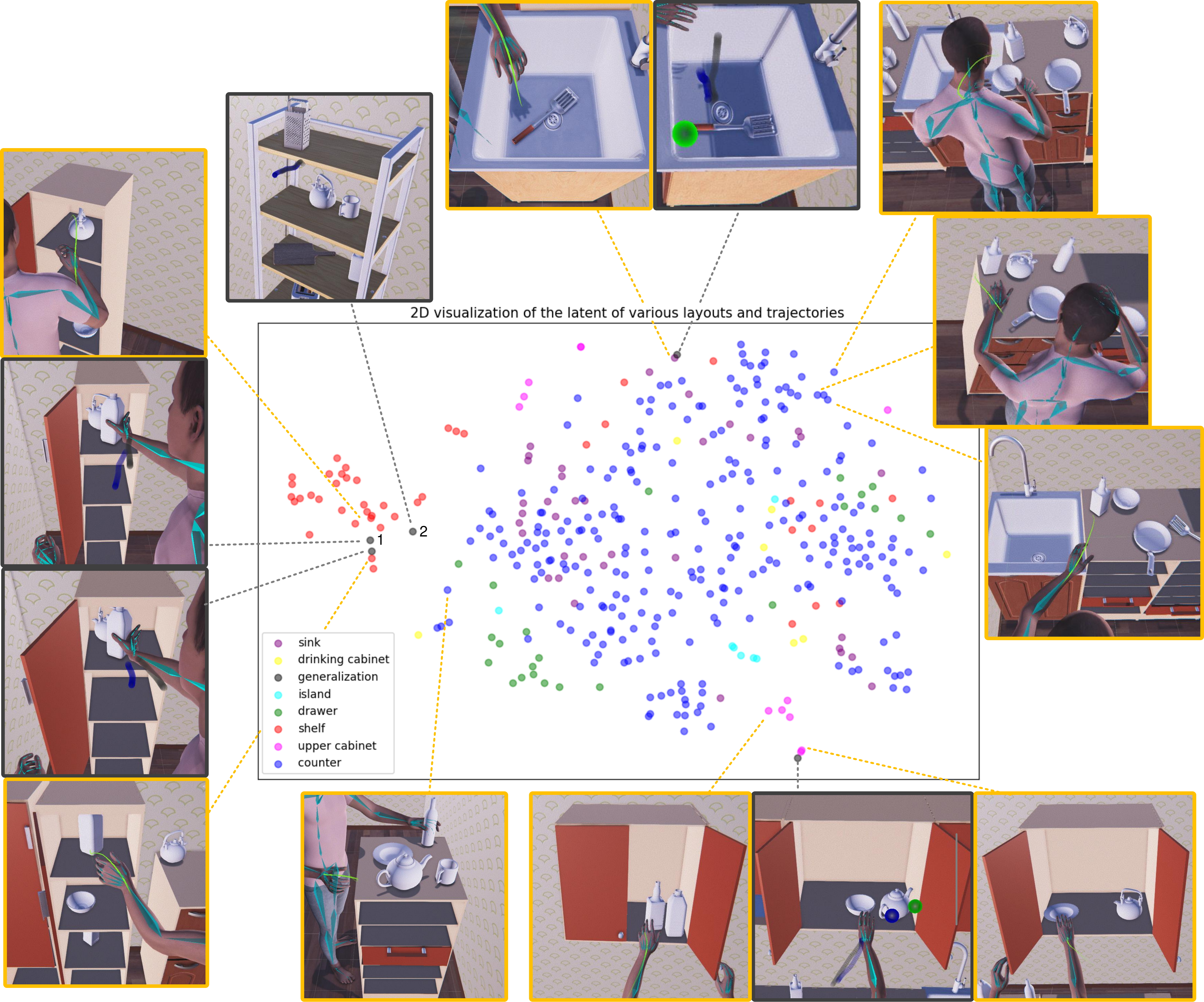}
		\caption{Visualization of the implicit neural representation of the layouts and approaching trajectories uniformly sampled from the training dataset of Sec.~\ref{sec:INTP}. We project the latent vector $\mbz$ of each 3-channel field $\mathbf{D}$ to 2D by t-SNE, discarding the condition labels $\mathbf{c}$. 
			The latent space automatically clusters different types of furniture, while similar layouts are continuously distributed.
			The projection of all the unseen layouts (cases in grey plots) is mapped to those of similar scenes (in orange).}
		\label{fig:latentDist}
	\end{figure}

	\begin{table}[]
		\caption{Comparison among SE(3) trajectory planners for the end effector. The spare safety distance to the nearest obstacle is averaged by frame among all the collision-free successful cases, counting the pieces of trajectories within 20cm around the touching/releasing position. We run our planner (INTPs) three times at each scene with different latent initializations sampled from Gaussian noise.}
		\label{tab:compTraj}
		\begin{tabular}{l|r|r|r}
			\hline
			\hline
			\multicolumn{4}{c}{Scene generalizations w/o reshaping} \\ \hline
			& unsmooth. $\downarrow$ & safety dist. $\uparrow$ & succ. rate $\uparrow$  \\ \hline
			cuRobo &     \textbf{4.07}     &    7.64     &  \textbf{100} \\   
			Diffusion      &    4.74      &   6.88 & 82.9 \\  \hline
            INTP-2Ch-$D_{t}$ & 4.21 $\pm$ 0.10  & 8.15 $\pm$ 0.11 & 97.6 \\
			INTP-2Ch-$D_{o'}$ & 4.32 $\pm$ 0.16  & 7.09 $\pm$ 0.19 & 91.7 \\
			INTP-3Channels     &     4.30 $\pm$ 0.09      &  \textbf{8.66 $\pm$ 0.13} & 98.8 \\ \hline
			\hline
			\multicolumn{4}{c}{Scene generalizations w/ reshaping \& novel furniture} \\ \hline
			cuRobo&     5.01     &    7.48     &  \textbf{96.4} \\  
			Diffusion      &    5.55      &   6.09 & 71.9 \\  \hline
            INTP-2Ch-$D_t$  & 5.26 $\pm$ 0.17  & 7.34 $\pm$ 0.19  & 89.9 \\
			INTP-2Ch-$D_{o'}$  & 5.58 $\pm$ 0.29  & 5.74 $\pm$ 0.23  & 78.6\\
			INTP-3Channels     &    \textbf{4.85  $\pm$ 0.20}       & \textbf{ 7.82 $\pm$ 0.15} & 95.8 \\ \hline
			\hline
		\end{tabular}
	\end{table}
	\subsubsection{Evaluation Under Scene Generalization}
	We further examine the generalization capability of our planner by applying it to unseen layouts and reshaped furniture.

	\paragraph{Trajectory Adaptation to Scenes}
	Although our planner was only trained on standard-sized furniture in our motion capture scenes, it learns the spatial relationships between the trajectories and the environment, and can produce trajectories in novel scenes that preserve such relationships.
	In~\figref{visGene}, we show snapshots of the pick-and-place motions on shelves, along with the adapted trajectories in blue and the predicted time-of-arrival field $\bm{D}_{toa}$ depicted in green. Under a standard-sized shelf with an ordinary placement (\figref{visGene}(a)), there is ample space for the right hand to reach the goal from the right side. When the shelf's size is narrowed to 80\% of the training size (see~\figref{visGene}(b)), the character instead reaches from the front and passes through the small space on the right to grasp the bottle. Further, after setting the bottle's position deeper into the narrow shelf (\figref{visGene}(c)), the character moves its wrist further to the left before reaching for the object. Upon adding another bottle (\figref{visGene}(d)), the system further adjusts the hand trajectory to avoid any collision. These experiments demonstrate that our designed three-channel fields of similar layouts can be well-mapped to a continuous latent space by the auto-decoder, enabling our planner to generalize to novel scenes.

    \paragraph{Latent Visualization}
	To further illustrate the generalization capability, we analyze the latent space of our auto-decoder.  Similar to how DeepSDF exhibits continuous latent representations for diverse shapes, our auto-decoder maps diverse scene–trajectory pairs to the latent space. We project the latent codes onto a plane using t-SNE~\cite{cieslak2020t} and visualize them in~\figref{latentDist}. 
	The projected scenes are well clustered, with unseen test scenes mapped adjacent to similar ones.
	This distribution reveals that even when training on scenes with various heterogeneous furniture together in one model, the auto-decoder structure effectively forms a continuous and smooth latent space for all types of scenes. This allows our runtime optimization to stably find the code that best suits the test environment and conditions, mapping it to a 6D trajectory within the distribution of plausible manipulations.
	
	\paragraph{Quantitative Evaluation}
	To quantitatively evaluate the generalization capability of different trajectory planners, we prepare an extended test dataset composed of our test data and those from indoor scene databases \footnote{https://assetstor.unity.com/packages/3d/props/interior/kitchen-props-pack-pbr-204679}\label{foot:scene}; featuring various obstacle shapes, layout differences, and placement variations. Similar to~\figref{visGene}, we synthesize unseen layouts by randomly shifting each obstacle object on/inside the involved furniture within 5$cm$, rotating them within $\pm90^\circ$, and re-scaling them within $0.8 \sim 1.25\times$. Also, we reshape the containers to new widths, depths, and heights ranging from 70\% to 150\% of their original dimensions, and remain the pickable cases.
	
	We compare the trajectory un-smoothness (measured by the mean acceleration), the remaining safety distance to obstacles from the hand and object, and the success rate among the results from our planner, the NVIDIA cuRobo planner, and the Diffusion Policy, as the examples presented in~\tabref{compTraj}. Our evaluation benchmark includes 168 densely packed layouts, which will be released alongside our Unity3D project.

	As shown in~\tabref{compTraj} part 2, our SE(3) trajectory planner achieves a high success rate over 96\%, with a 4.5\% higher safety distance than the cuRobo planner and 28.4\% higher than the Diffusion Policy, along with significantly better motion smoothness. These results validate the benefits of our planner's strong generalization capability and more human-like pick-and-place trajectories. The success rate of cuRobo also decreases when the scene is highly cluttered and requires more adjustments to navigate through narrow regions. In these cases, employing a highly precise geometric approximation in cuRobo fails to converge to a feasible solution, while using a coarser geometric approximation relaxes constraints excessively, leading to collision-prone policies.

\paragraph{Qualitative Evaluation}
Our generalization benchmark reveals that the deformation of the inference field enables our planner to perform robustly in unseen, cluttered environments. In contrast, the Diffusion Policy, also as a data-driven planner, fails in narrow scenes due to collisions, as shown in cases a-3 to c-3 in~\figref{visCompareGene}. In these scenarios, the Diffusion model's planning remains similar to the easier cases predominantly featured in the training data, which involve only ordinary-sized cabinets. These shortcomings become more pronounced when the layout and obstacle shapes significantly differ from the dataset (see supplementary video). Although this model could be improved by capturing more demonstrations from cabinets with greater diversity in shapes and configurations, the associated motion capture cost increases rapidly.
	
	\subsubsection{Ablation Study on Distance Field Channels}
	 We evaluate the effectiveness of our three-field setting by comparing it with a simplified two-channel auto-decoder. In these alternative models, we only keep the goal-reaching distance $D_t$ (as INTP-2Ch-$D_t$) or we merge the $D_t$ together with the obstacle distance $D_o$ into one channel $D_{o'}$ (as INTP-2Ch-$D_{o'}$) by the unsigned/signed distance to the nearest obstacle/target object surface, respectively. As shown in~\tabref{compTraj}, the performance decreases significantly when using only two channels. This result confirms that having separate and complete channels for $D_t$ and $D_o$ is beneficial for high-quality trajectory predictions. Specifically, the $D_t$ field away from the object helps to direct the hand to approach, while the $D_o$ field contributes to collision avoidance.
    
    Moreover, due to the parallel optimization at runtime, reducing the number of output feature channels does not lead to a reduction in time or space complexity. Therefore, it is advantageous to utilize all accessible runtime information by distributing it across multiple channels. This approach allows our network to better infer the $D_{toa}$ field through our proposed structure.

	\begin{figure*}[t]
		\includegraphics[width=0.8\linewidth]{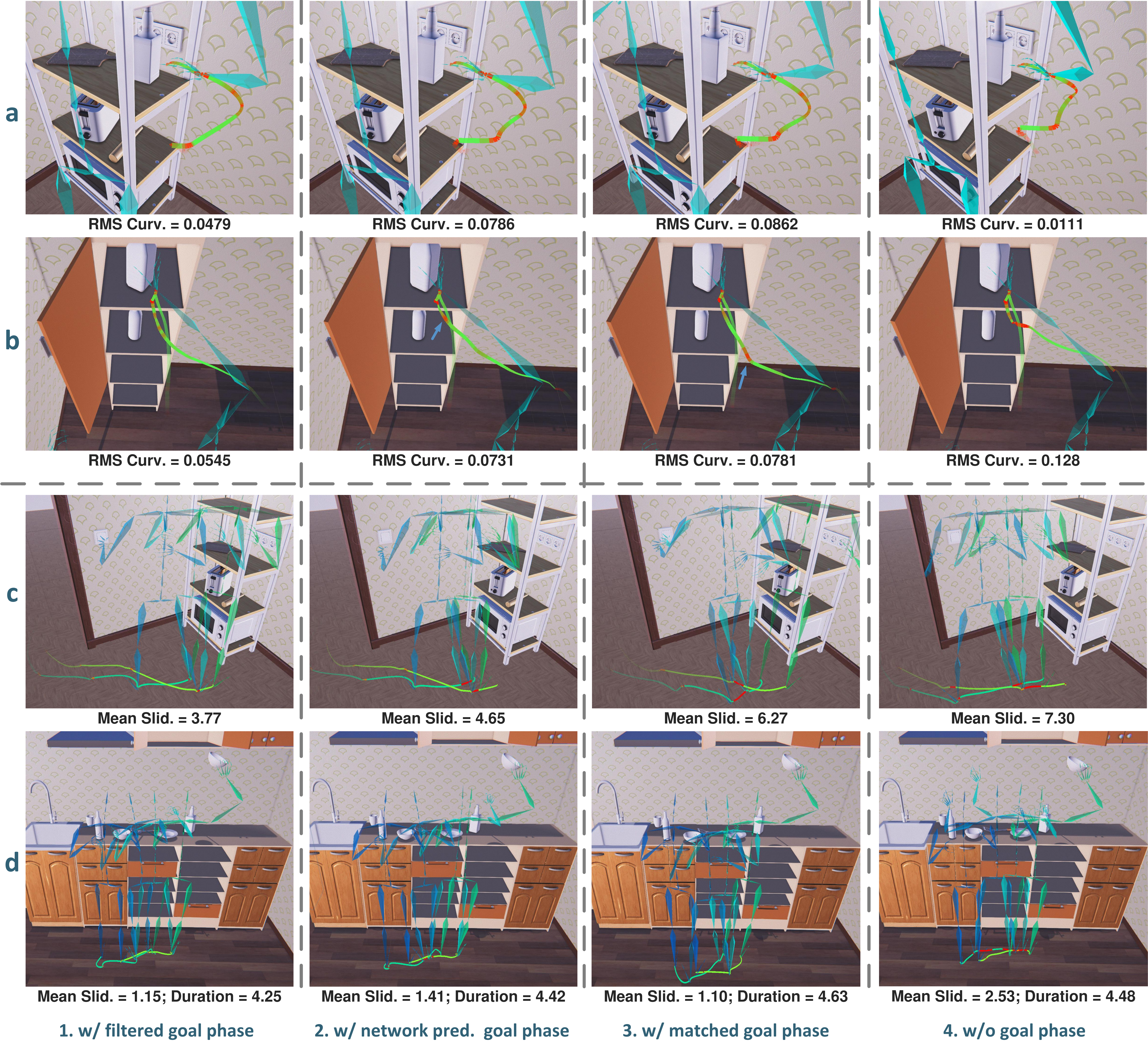}
		\caption{Ablation study for our DeepPhase interaction controller in typical full-body interactions. We visualize the hand trace for the target approaching at (a,b), where the trajectory pieces with high curvature are highlighted by red heatmaps. In line (c), we show the bipedal trace of a pivot turning case, and line (d) shows a side-stepping case, where the skeletons from green to blue are from sequentially generated key-frames where both feet are on the ground, with red curves highlighting the sliding pieces. We annotate each approaching trajectory with its RMS curvature in $cm^{-1}$, the mean sliding during stepping in $cm/s$, and the duration time in seconds.}
		\label{fig:visAblation}
	\end{figure*}
	
	\subsection{Evaluation of the Pick-and-Place Motions}
	\label{sec:ablationExp}
	We show long-term interaction examples where the user produces pick-and-place motions through mouse-clicking objects in various environments.  
	As the user clicks an object or a target location, the character automatically walks towards the object/destination and picks/places the object (see~\figref{teaser}). The character can adapt appropriate policies to achieve the goal task with its unoccupied hand. Even under highly cluttered scenes, it executes a series of actions smoothly to remove obstacles before picking it up. Readers are referred to the supplementary video for detailed demonstrations.
	By testing various long-term interactions as demonstrated, we further evaluate the coordinated body motion quantitatively/qualitatively by conducting an ablation study on the goal-driven controller.

	\subsubsection{Ablation Study of the Deep Phase Interaction Controller}
	We evaluate the effectiveness of our goal-driven character controller by examining 3 variants of our DeepPhase interaction controller, including a model that drives the character solely based on the goal key joints (w/o any goal phase), a model that only uses matched goal phase features obtained from the dataset (w/o predicted goal phase), and a model where the network is only fed with goal phase features auto-regressively predicted by itself
	(w/o matched goal phase).
    In addition to testing in the general scene described earlier (see Footnote~\ref{foot:scene}), we use another test scene from our dataset composed of 20 lightweight objects and 9 pieces of furniture. A long motion sequence over 3 minutes is constructed by the user through sequential clicks on objects and furniture. We compute the smoothness, foot skating artifacts, and the Fr\'{e}chet distance to the dataset (see~\tabref{ablation}). The results show that the use of the goal phase improves the accuracy of all measures.  

    \paragraph{Evaluation on Full-body Coordination}
	As the phase feature is computed based on the full body motion, it helps to coordinate all parts of the body to produce realistic movements.
    
		When no goal phase feature is given, the legs fail to produce effective stepping to let the key joints track their goals, resulting in larger foot skating (see \tabref{ablation}). This is evident in complex locomotion like pivot turning (see \figref{visAblation}(c-4)). Also, the upper body can struggle to recover from challenging previous poses without a goal cue, as shown in~\figref{visAblation}(d-4), where the uncoordinated body causes increased foot sliding.
        
		When the matched goal phase is used instead of that predicted by the network, the system suffers from poor guidance to the goal. 
        As the matched goal phase corresponds to sparse keyframes, it cannot guide the character to conduct efficient stepping for reaching a state that is spatio-temporally apart (see~\figref{visAblation}(c-3)).   
    Also, the deviation of the current state and the matched state leads to biased directional guidance, requiring the character to take extra steps/time for reaching the goal (see~\figref{visAblation}(d-3)).
        
		Although the locomotion is natural and smooth when guided by the network-estimated goal phase, relying solely on network prediction and regression can omit some high-frequency signals, occasionally causing worse sliding performance compared to that guided by the matched goal prior (see~\figref{visAblation}(d-2)).
        
        Since the matched goal phase provides specific goal states from similar cases, our filtered goal phase corrects the matched phase to fit the test-scene key joint goals. By combining the matched goal phase with the network prediction using the Kalman filter, we simultaneously maintain high-frequency signals and appropriately guide the locomotion to efficiently transition toward the next interaction task (see~\figref{visAblation}(c-1, d-1)).

    \paragraph{Evaluation on Hand Control}
    
    Furthermore, the trajectory of the end-effector exhibits higher motion realism and smoothness, accompanied by body coordination when using the filtered goal phase. Without the goal cue in the frequency domain, tracking only the three key joint goals spatially leads the model to focus excessively on tracking the hip and the non-manipulating hand goals, especially when the manipulation takes place away from the torso. This causes large deviations when approaching the hand goals (see~\figref{visAblation}(b-4)) and results in an unsmooth trajectory of the manipulating hand.

Even provided with the matched goal phase priors, the system can produce unsmooth wrist motions when the matched prior suddenly switches from left-hand grasping to right-hand grasping—for example, when the main hand starts to reach out for the target object after the auxiliary hand finishes opening the cabinet, as shown in~\figref{visAblation}(b-3), and significant direction adjustments can also occur (see~\figref{visAblation}(a-3)).

While the approaching motion toward the goal phase estimated directly by the MoE network slightly improves curvature results (see~\figref{visAblation}(a-2, b-2)), it still suffers from discrete switches of the goal phase. When moving the hand goal spatially along the planned approaching trajectory, the predicted goal phase updates rapidly to catch up with the goal. In case (b-2), the arm immediately stretches out to reach the target object even before the character is close enough, producing an awkward motion due to overstretching.

In contrast, the Kalman filter smoothly transitions the goal phase by combining the goal prediction with the matched grasping motion and the system's dynamic recurrence. This results in a smooth hand trajectory well aligned with active stepping toward the target to assist the grasping process (see~\figref{visAblation}(a-1, b-1)). Additionally, it performs highly realistic small lower-body stepping during left-hand manipulation before the right-hand task, as shown in both cases (a-1, b-1). This slight body adjustment contributes to conducting the ongoing manipulation with high smoothness.

\begin{figure}[t]
    \includegraphics[width=\columnwidth]{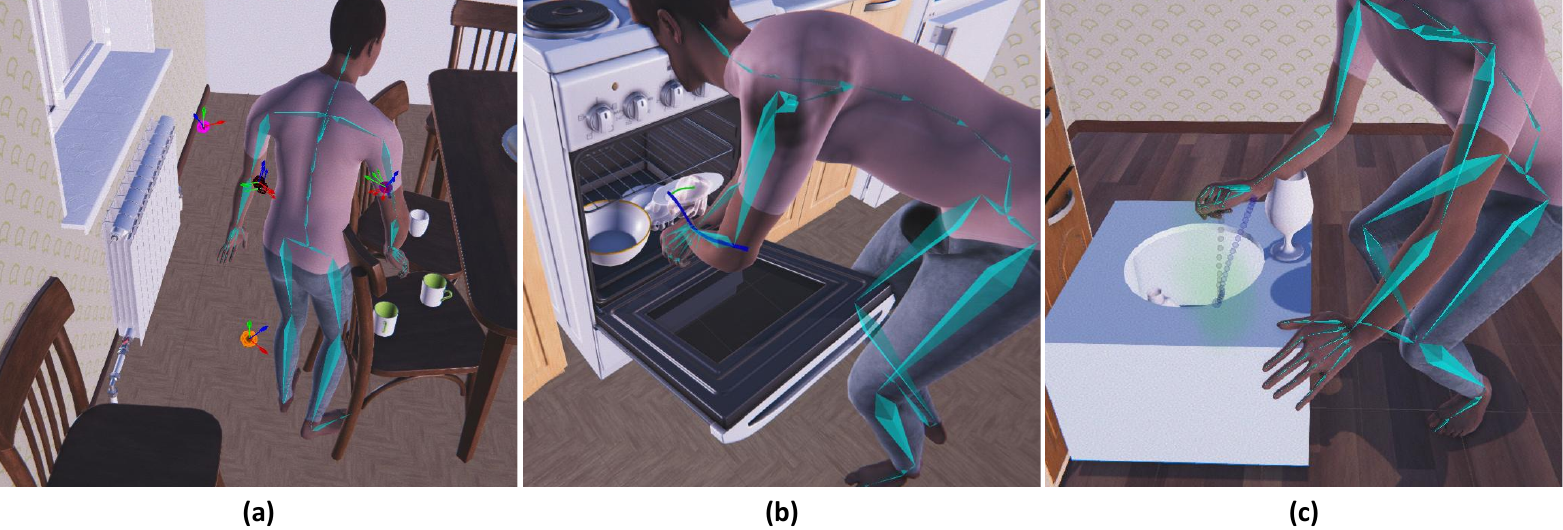}
    \caption{Limited performance on generating coordinate full-body collision-free interaction under novel scenes.}
    \label{fig:limitation}
\end{figure}

	\section{Discussion and Future Works}
     
	In this work, we demonstrate the effectiveness of our implicit neural trajectory planner in generating realistic pick-and-place guidance with strong generalization across various layouts. Combined with a robust goal-driven controller that simultaneously estimates the tracking state in the frequency domain, our system drives the character to perform complex interactions guided by the goals of three key joints. This capability is further enhanced by our bimanual scheduler. Together, these techniques are hierarchically integrated to form the first systematic method for long-term interaction synthesis in cluttered environments. The proposed framework can be extended in several ways to tackle even more complex scenarios.
 
	Currently, our neural implicit trajectory planner is limited to pick-and-place motions. It can potentially be enhanced to handle more complex manipulation tasks, such as cooking or assembling/disassembling mechanical objects. For such purposes, it will be interesting to look into generalizing vector fields that have curls, or those that dynamically change according to the configuration of the body or the objects. 

	The containers we handle are limited to simple cabinets, drawers, and shelves, with fixed procedures for opening and closing them. Although this simplifies the scheduling of the planner, it cannot generalize to arbitrary containers with different types of doors, lids, and openers. Establishing a framework that can learn the structures of various container types and the procedures required to open them and retrieve contents would be an interesting research direction.

    The characters are currently controlled by mouse-clicking, which is intuitive but requires some effort, as the viewpoint also may need to be managed by the user. It will be interesting to look into applying LLMs to control the character by using language as an interface.

    \paragraph{Limitations}
    
   Despite the strengths exhibited in our evaluations, our pipeline has certain limitations as follows.

    \paragraph{Navigation-Induced Collisions} Our full-body navigation towards the goal is currently based on a 2D path planner and navigation matching. Thus, the character cannot well avoid obstacles that require the body to bend down, tilt the torso to the side, or walk over. 
    As a result, the reaching motion generated through tracking a sequence of keyjoint goals is not guaranteed to be collision-free between the goal waypoints. It is difficult to resolve collisions in an online manner, as this can degrade the motion coordination and smoothness (see~\figref{limitation}(a) and the supplementary video). One possibility is to incorporate sensors that evaluate the distances to 3D objects for the 2D path planning and to use such sensors for motion synthesis during runtime~\cite{holden2017phase,starke2019neural}. 
    
    \paragraph{Generalization Gaps to Unseen Furniture} Performing collision-free and coordinated human interaction in arbitrary cluttered environments presents significant challenges.
    Our current system exhibits limitations in generalization to unseen furniture that is significantly different from that in our dataset (see~\figref{limitation}).     
    For full-body motion, shifting the 3-keyjoint goal transformations for collision-resolving (described in~\secref{MotionGoalSynthesis}) can lead to uncoordinated motion. E.g., in~\figref{limitation}(b), the increased hand-hip distance imposed by the oven door results in an unnatural full-body motion with foot sliding. Additionally, on the manipulating hand, collisions sometimes occurred due to the failed trajectory planning, when it cannot effectively guide the manipulating hand to pass through surrounding furniture obstacles. For instance, in~\figref{limitation}(c), an ideal trajectory should guide the hand to reach into and depart from the box's opening while avoiding the wineglass.

    However, these issues can be alleviated through computing an optimal goal-reaching pose adaptive to diverse scenes, by leveraging prior knowledge extracted from a vast video dataset of human interactions. For example, instead of approaching from the front, as shown in~\figref{limitation}(b,c), the character could potentially reach the target more closely from less obstructed sides, thereby mitigating the conflict between humanoid collision avoidance and realistic grasping. To achieve this, we could first learn a probabilistic 2D pose prior from extensive, unlabeled interaction videos, which acts as a realism evaluator. A plausible goal pose could then be found by programming in the canonical space, with the objective of maximizing the score aggregated from this evaluator across multiple 2D projections, subject to the test-time collision constraints.
     
\begin{acks}
This project was partially funded by Meta. It was also partially funded by the Research Grants Council of Hong Kong (Ref: 17210222), and by the Innovation and Technology Commission of the HKSAR Government under the ITSP-Platform grant (Ref: ITS/335/23FP) and the InnoHK initiative (TransGP project). Part of the research was conducted in the JC STEM Lab of Robotics for Soft Materials, funded by The Hong Kong Jockey Club Charities Trust.
\end{acks}

	\bibliographystyle{ACM-Reference-Format}
	\bibliography{sample-bibliography}

%% file: samplebody-journals-Supp.tex
\renewcommand{\thesection}{S\arabic{section}}
\renewcommand{\thetable}{S\arabic{table}}%
\renewcommand{\thefigure}{S\arabic{figure}}%

\begin{figure}[]
    \includegraphics[width=\columnwidth]{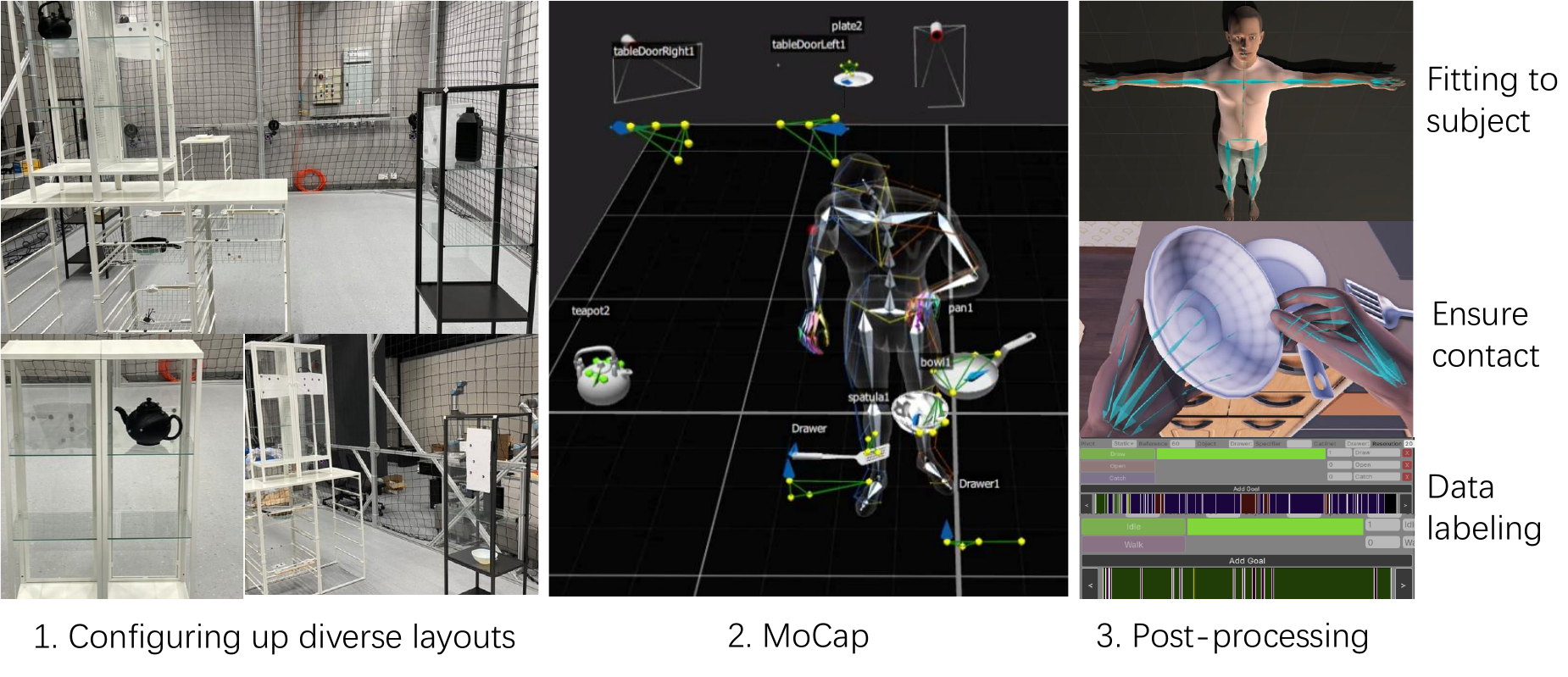}
    \caption{A brief procedure for capturing our interaction dataset.}
    \label{fig:MoCap}
\end{figure}

\section{Motion Capture Procedure}
    We illustrate our motion capture procedure in~\figref{MoCap}. For each motion sequence, we first set up an indoor scene composed of various pieces of furniture arranged in adjusted layouts. We then perform a long motion sequence after ensuring that all pickable objects inside are visible, as described in §7. After acquiring all the raw marker sequence data, we optimize the blendshape of an SMPL model and the Vicon hand skeleton across all sequences, and solve bone transformations with precise hand-object contact while also resolving self-collisions. After exporting the processed motion sequences to Unity3D, we label each sequence with its actions, goal keyframes, contacts, and phases as detailed in §7.

\begin{table}[]
	\centering
	\caption{Input/output features of each module for motion control. The feature dimensions are listed in the order of channel numbers (e.g., joint numbers, DeepPhase channels), length of each feature vector, and the frame numbers.  All SE(3) transformations are projected to $\mathbb{R}^9$ by dual-axes~\cite{xiang2021eliminating}. The motion prediction network auto-regressively updates the current-centric trajectories using 13 keyframes extracted from $[-1s,1s]$ by the guidance of 7 keyframes (1s time window) goals extracted since the goal frame.}
	\label{tab:features}
	\begin{tabular}{c|c}

	    \hline\hline
		\multicolumn{2}{c}{\textbf{Goal Coordination by a State Machine (§6)}} \\
		\hline
		
		\textit{Input}  & \textit{Feat. Dim.}\\
		\hline
		2D path for root navigation & 2$\times len.$\\
		Keyjoint transformations for navigation (§6.2)& 3$\times$9 \\
		Planned wrist trajectory $\in SE(3)$& 9$\times len.$ \\
		Keyjoint goals from bimanual matching (§6.3) & 3$\times$9$\times$270 \\
		All colliders in the environment & - \\
		\hline
		\textit{Output}  & \textit{Feat. Dim.}\\ \hline
		Goal keyjoint transformations $\in SE(3)$ & 3$\times$9$\times$7	\\
		Matched DeepPhase goal vectors $\in [0,1]$ & 8$\times$2$\times$13 \\

	    \hline\hline
		\multicolumn{2}{c}{\textbf{Motion Prediction Network (§5.1)}}\\
		\hline
		\textit{Gating Net. Input}  & \textit{Feat. Dim.}\\
		\hline
		
		Current hand contact labels $\in \{0,1\}$ & 2 \\
		Current keyjoint transformations $\in SE(3)$ & 3$\times$9\\
		Current DeepPhase vectors $\in [0,1]$ & 8$\times$2$\times$13 \\
		Goal keyjoint transformations $\in SE(3)$ & 3$\times$9$\times$7 \\
		Goal hip and hand action labels $\in [0,1]$ & (2+3+3)$\times$7 \\
		DeepPhase goal vectors $\in [0,1]$ & 8$\times$2$\times$13 \\
		
		\hline
		\textit{Main Net. Input}  & \textit{Feat. Dim.}\\
		\hline
		Current root trajectory $\in SE(2)$ + vel. $\in \mathbb{R}^+
$ & 8$\times$13 \\
		Current keyjoint trajectories $\in SE(3)$ + vel. $\in \mathbb{R}^+$ & 3$\times$12$\times$13\\
		Root-centric pose $\in SE(3)$ + joint vel. $\in \mathbb{R}^+
$ & 29$\times$12 \\	
		\hline
		\textit{Main Net. Output}  & \textit{Feat. Dim.}\\ \hline
		Next ego-centric root traj. $\in SE(2)$ + vel. $\in \mathbb{R}^+
$  & 8$\times$7 \\
Next goal-centric keyjoint traj. $\in SE(3)$ & 3$\times$9$\times$7 \\
		Next ego-centric keyjoint traj. $\in SE(3)$ + vel. $\in \mathbb{R}^+
$  & 3$\times$12$\times$7 \\
Next goal-centric root traj. $\in SE(2)$ & 6$\times$7 \\
		
		Next root-centric pose $\in SE(3)$ + joint vel. $\in \mathbb{R}^+
$ &  29$\times$12 \\
		Next pose relative to the root 1s ahead $\in SE(3)$ &  29$\times$9 \\
		Next hip and hand action labels $\in [0,1]$ & (2+3+3)$\times$7 \\
		Next foot and hand contact values $\in [0,1]$ & 4 \\
		DeepPhase updates on phase, amp., and freq. & 8$\times$4$\times$7 \\
		Predicted goal keyjoint transformations $\in SE(3)$ & 3$\times$9$\times$7 \\
		\hline\hline
		
		\multicolumn{2}{c}{\textbf{Kalman Filter  (§5.2)}} \\
		\hline
		
		\textit{Input}  & \textit{Feat. Dim.}\\
		\hline
		Current DeepPhase goal vectors $\in [0,1]$& 8$\times$2$\times$13 \\
		Matched DeepPhase goal vectors $\in [0,1]$& 8$\times$2$\times$13 \\
		Predicted DeepPhase goal vectors $\in [0,1]$& 8$\times$2$\times$13 \\
		Predicted ego/goal-centric keyjoint traj. $\in SE(3)$ & 2$\times$3$\times$9$\times$7 \\
		\hline
		\textit{Output}  & \textit{Feat. Dim.}\\ \hline
		Next DeepPhase goal phase, amp., and freq.  & 8$\times$4$\times$13 \\
		\hline\hline
	\end{tabular}
\end{table}

\section{Implementation details in Motion control}
In~\tabref{features}, we elaborate on the input/output features of the modules in the DeepPhase interaction controller described in §5 and the goal coordination in §6.3. After running the three modules in each auto-regressive loop, we post-process the predicted character pose by IK and bi-directionally blend the keyjoint trajectory predictions. For manipulation, after the system computes the pose, we apply IK to the arms, informed by the collision-avoidance poses computed by an RRT* search. For the lower body, we apply CCD-IK to the feet using the predicted foot contact values, following~\cite{starke2019neural}.

\begin{figure}[]
    \includegraphics[width=\columnwidth]{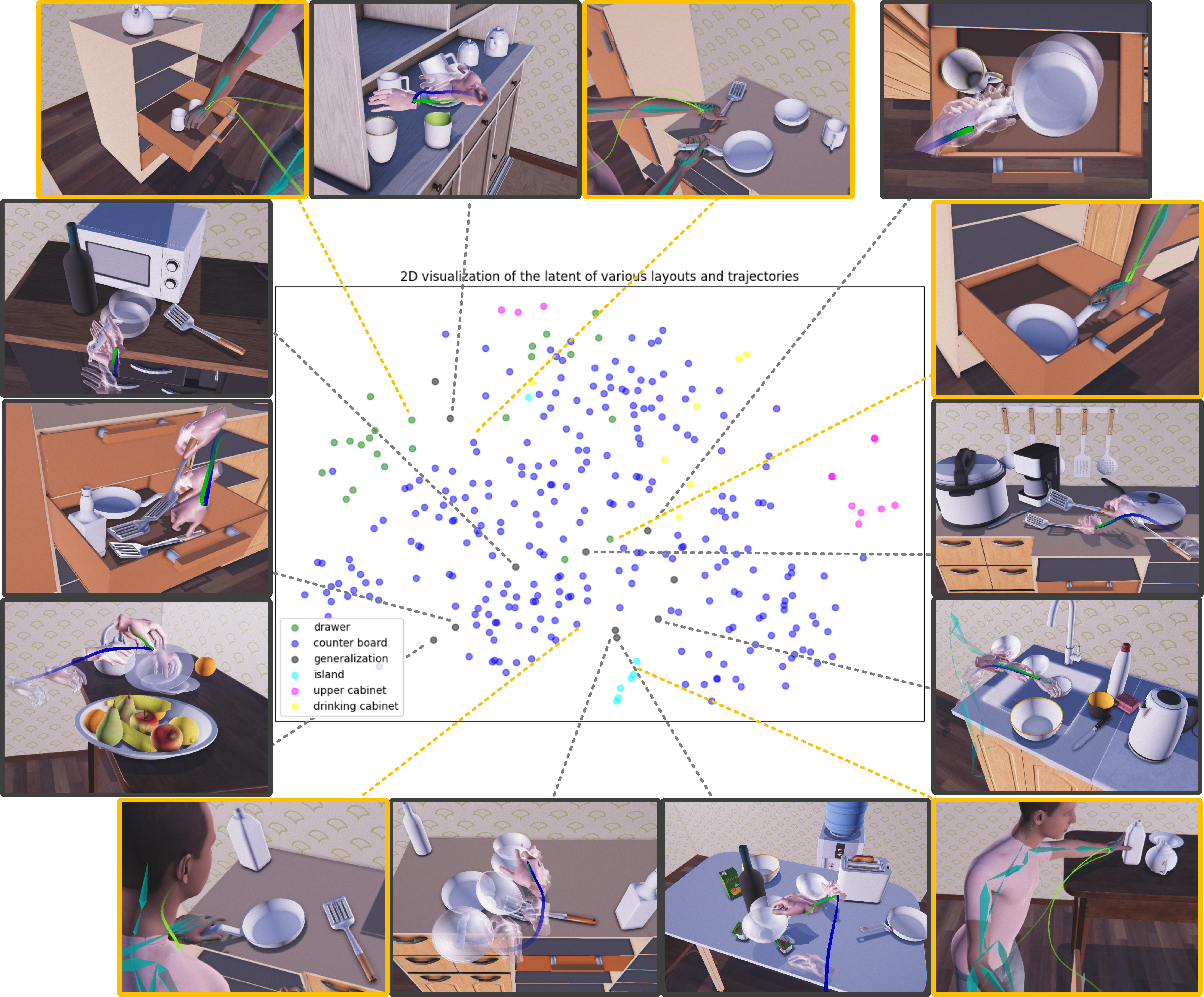}
    \caption{Visualization of more test cases under scene generalization, applying our implicit neural representation for unseen table/cupboard/drawer-board layouts and trajectories (in grey). The data distribution corresponds to the central part of Fig. 11.}
    \label{fig:latentDist1}
\end{figure}

\section{Latent Visualization of Layouts on Boards}
 An extended version of Fig.~11 in the main text is shown in~\figref{latentDist1} here. More cases corresponding to the central parts of Fig.11 are added, consisting mainly of board-top layouts from drawers, cupboards, and island tables. The scenes in the grey boxes are randomly generated. The projection visualizes that they are located near similar layouts from the training set shown in orange boxes, all distributed within a continuous latent distribution.

%% file: sample-acmtog-SIGGRAPH-submission.bbl

\begin{thebibliography}{2}


\ifx \showCODEN    \undefined \def \showCODEN     #1{\unskip}     \fi
\ifx \showDOI      \undefined \def \showDOI       #1{#1}\fi
\ifx \showISBNx    \undefined \def \showISBNx     #1{\unskip}     \fi
\ifx \showISBNxiii \undefined \def \showISBNxiii  #1{\unskip}     \fi
\ifx \showISSN     \undefined \def \showISSN      #1{\unskip}     \fi
\ifx \showLCCN     \undefined \def \showLCCN      #1{\unskip}     \fi
\ifx \shownote     \undefined \def \shownote      #1{#1}          \fi
\ifx \showarticletitle \undefined \def \showarticletitle #1{#1}   \fi
\ifx \showURL      \undefined \def \showURL       {\relax}        \fi
\providecommand\bibfield[2]{#2}
\providecommand\bibinfo[2]{#2}
\providecommand\natexlab[1]{#1}
\providecommand\showeprint[2][]{arXiv:#2}

\bibitem[Starke et~al\mbox{.}(2019)]%
        {starke2019neural}
\bibfield{author}{\bibinfo{person}{Sebastian Starke}, \bibinfo{person}{He Zhang}, \bibinfo{person}{Taku Komura}, {and} \bibinfo{person}{Jun Saito}.} \bibinfo{year}{2019}\natexlab{}.
\newblock \showarticletitle{Neural state machine for character-scene interactions.}
\newblock \bibinfo{journal}{\emph{ACM Trans. Graph.}} \bibinfo{volume}{38}, \bibinfo{number}{6} (\bibinfo{year}{2019}), \bibinfo{pages}{209--1}.
\newblock


\bibitem[Xiang(2021)]%
        {xiang2021eliminating}
\bibfield{author}{\bibinfo{person}{Sitao Xiang}.} \bibinfo{year}{2021}\natexlab{}.
\newblock \showarticletitle{Eliminating topological errors in neural network rotation estimation using self-selecting ensembles}.
\newblock \bibinfo{journal}{\emph{ACM Transactions on Graphics (TOG)}} \bibinfo{volume}{40}, \bibinfo{number}{4} (\bibinfo{year}{2021}), \bibinfo{pages}{1--21}.
\newblock


\end{thebibliography}



\begin{thebibliography}{65}


\ifx \showCODEN    \undefined \def \showCODEN     #1{\unskip}     \fi
\ifx \showDOI      \undefined \def \showDOI       #1{#1}\fi
\ifx \showISBNx    \undefined \def \showISBNx     #1{\unskip}     \fi
\ifx \showISBNxiii \undefined \def \showISBNxiii  #1{\unskip}     \fi
\ifx \showISSN     \undefined \def \showISSN      #1{\unskip}     \fi
\ifx \showLCCN     \undefined \def \showLCCN      #1{\unskip}     \fi
\ifx \shownote     \undefined \def \shownote      #1{#1}          \fi
\ifx \showarticletitle \undefined \def \showarticletitle #1{#1}   \fi
\ifx \showURL      \undefined \def \showURL       {\relax}        \fi
\providecommand\bibfield[2]{#2}
\providecommand\bibinfo[2]{#2}
\providecommand\natexlab[1]{#1}
\providecommand\showeprint[2][]{arXiv:#2}

\bibitem[Araujo et~al\mbox{.}(2023)]%
        {araujo2023circle}
\bibfield{author}{\bibinfo{person}{Joao~Pedro Araujo}, \bibinfo{person}{Jiaman Li}, \bibinfo{person}{Karthik Vetrivel}, \bibinfo{person}{Rishi Agarwal}, \bibinfo{person}{Deepak Gopinath}, \bibinfo{person}{Jiajun Wu}, \bibinfo{person}{Alexander Clegg}, {and} \bibinfo{person}{C.~Karen Liu}.} \bibinfo{year}{2023}\natexlab{}.
\newblock \showarticletitle{CIRCLE: Capture in Rich Contextual Environments}. In \bibinfo{booktitle}{\emph{Proceedings of the IEEE/CVF conference on computer vision and pattern recognition}}.
\newblock


\bibitem[Bae et~al\mbox{.}(2023)]%
        {bae2023pmp}
\bibfield{author}{\bibinfo{person}{Jinseok Bae}, \bibinfo{person}{Jungdam Won}, \bibinfo{person}{Donggeun Lim}, \bibinfo{person}{Cheol-Hui Min}, {and} \bibinfo{person}{Young~Min Kim}.} \bibinfo{year}{2023}\natexlab{}.
\newblock \showarticletitle{Pmp: Learning to physically interact with environments using part-wise motion priors}. In \bibinfo{booktitle}{\emph{ACM SIGGRAPH 2023 Conference Proceedings}}. \bibinfo{pages}{1--10}.
\newblock


\bibitem[Bhatnagar et~al\mbox{.}(2022)]%
        {bhatnagar2022behave}
\bibfield{author}{\bibinfo{person}{Bharat~Lal Bhatnagar}, \bibinfo{person}{Xianghui Xie}, \bibinfo{person}{Ilya~A Petrov}, \bibinfo{person}{Cristian Sminchisescu}, \bibinfo{person}{Christian Theobalt}, {and} \bibinfo{person}{Gerard Pons-Moll}.} \bibinfo{year}{2022}\natexlab{}.
\newblock \showarticletitle{Behave: Dataset and method for tracking human object interactions}. In \bibinfo{booktitle}{\emph{Proceedings of the IEEE/CVF Conference on Computer Vision and Pattern Recognition}}. \bibinfo{pages}{15935--15946}.
\newblock


\bibitem[Braun et~al\mbox{.}(2024)]%
        {braun2023physically}
\bibfield{author}{\bibinfo{person}{Jona Braun}, \bibinfo{person}{Sammy Christen}, \bibinfo{person}{Muhammed Kocabas}, \bibinfo{person}{Emre Aksan}, {and} \bibinfo{person}{Otmar Hilliges}.} \bibinfo{year}{2024}\natexlab{}.
\newblock \showarticletitle{Physically Plausible Full-Body Hand-Object Interaction Synthesis}. In \bibinfo{booktitle}{\emph{International Conference on 3D Vision (3DV)}}.
\newblock


\bibitem[Camps et~al\mbox{.}(2022)]%
        {camps2022learning}
\bibfield{author}{\bibinfo{person}{Gadiel~Sznaier Camps}, \bibinfo{person}{Robert Dyro}, \bibinfo{person}{Marco Pavone}, {and} \bibinfo{person}{Mac Schwager}.} \bibinfo{year}{2022}\natexlab{}.
\newblock \showarticletitle{Learning Deep SDF Maps Online for Robot Navigation and Exploration}.
\newblock \bibinfo{journal}{\emph{arXiv preprint arXiv:2207.10782}} (\bibinfo{year}{2022}).
\newblock


\bibitem[Cen et~al\mbox{.}(2024)]%
        {cen2024generating}
\bibfield{author}{\bibinfo{person}{Zhi Cen}, \bibinfo{person}{Huaijin Pi}, \bibinfo{person}{Sida Peng}, \bibinfo{person}{Zehong Shen}, \bibinfo{person}{Minghui Yang}, \bibinfo{person}{Shuai Zhu}, \bibinfo{person}{Hujun Bao}, {and} \bibinfo{person}{Xiaowei Zhou}.} \bibinfo{year}{2024}\natexlab{}.
\newblock \showarticletitle{Generating Human Motion in 3D Scenes from Text Descriptions}. In \bibinfo{booktitle}{\emph{Proceedings of the IEEE/CVF Conference on Computer Vision and Pattern Recognition}}. \bibinfo{pages}{1855--1866}.
\newblock


\bibitem[Chen et~al\mbox{.}(2022)]%
        {chen2022neural}
\bibfield{author}{\bibinfo{person}{Yun-Chun Chen}, \bibinfo{person}{Adithyavairavan Murali}, \bibinfo{person}{Balakumar Sundaralingam}, \bibinfo{person}{Wei Yang}, \bibinfo{person}{Animesh Garg}, {and} \bibinfo{person}{Dieter Fox}.} \bibinfo{year}{2022}\natexlab{}.
\newblock \showarticletitle{Neural motion fields: Encoding grasp trajectories as implicit value functions}.
\newblock \bibinfo{journal}{\emph{Robotics Science and Systems (RSS) Workshop on Implicit Representations for Robotic Manipulation}} (\bibinfo{year}{2022}).
\newblock


\bibitem[Chi et~al\mbox{.}(2023)]%
        {chi2023diffusion}
\bibfield{author}{\bibinfo{person}{Cheng Chi}, \bibinfo{person}{Zhenjia Xu}, \bibinfo{person}{Siyuan Feng}, \bibinfo{person}{Eric Cousineau}, \bibinfo{person}{Yilun Du}, \bibinfo{person}{Benjamin Burchfiel}, \bibinfo{person}{Russ Tedrake}, {and} \bibinfo{person}{Shuran Song}.} \bibinfo{year}{2023}\natexlab{}.
\newblock \showarticletitle{Diffusion policy: Visuomotor policy learning via action diffusion}.
\newblock \bibinfo{journal}{\emph{The International Journal of Robotics Research}} (\bibinfo{year}{2023}), \bibinfo{pages}{02783649241273668}.
\newblock


\bibitem[Christen et~al\mbox{.}(2022)]%
        {christen2022d}
\bibfield{author}{\bibinfo{person}{Sammy Christen}, \bibinfo{person}{Muhammed Kocabas}, \bibinfo{person}{Emre Aksan}, \bibinfo{person}{Jemin Hwangbo}, \bibinfo{person}{Jie Song}, {and} \bibinfo{person}{Otmar Hilliges}.} \bibinfo{year}{2022}\natexlab{}.
\newblock \showarticletitle{D-grasp: Physically plausible dynamic grasp synthesis for hand-object interactions}. In \bibinfo{booktitle}{\emph{Proceedings of the IEEE/CVF Conference on Computer Vision and Pattern Recognition}}. \bibinfo{pages}{20577--20586}.
\newblock


\bibitem[Cieslak et~al\mbox{.}(2020)]%
        {cieslak2020t}
\bibfield{author}{\bibinfo{person}{Matthew~C Cieslak}, \bibinfo{person}{Ann~M Castelfranco}, \bibinfo{person}{Vittoria Roncalli}, \bibinfo{person}{Petra~H Lenz}, {and} \bibinfo{person}{Daniel~K Hartline}.} \bibinfo{year}{2020}\natexlab{}.
\newblock \showarticletitle{t-Distributed Stochastic Neighbor Embedding (t-SNE): A tool for eco-physiological transcriptomic analysis}.
\newblock \bibinfo{journal}{\emph{Marine genomics}}  \bibinfo{volume}{51} (\bibinfo{year}{2020}), \bibinfo{pages}{100723}.
\newblock


\bibitem[Fan et~al\mbox{.}(2023)]%
        {fan2023arctic}
\bibfield{author}{\bibinfo{person}{Zicong Fan}, \bibinfo{person}{Omid Taheri}, \bibinfo{person}{Dimitrios Tzionas}, \bibinfo{person}{Muhammed Kocabas}, \bibinfo{person}{Manuel Kaufmann}, \bibinfo{person}{Michael~J. Black}, {and} \bibinfo{person}{Otmar Hilliges}.} \bibinfo{year}{2023}\natexlab{}.
\newblock \showarticletitle{{ARCTIC}: A Dataset for Dexterous Bimanual Hand-Object Manipulation}. In \bibinfo{booktitle}{\emph{Proceedings IEEE Conference on Computer Vision and Pattern Recognition (CVPR)}}.
\newblock


\bibitem[Ghorbani and Black(2021)]%
        {SOMA:ICCV:2021}
\bibfield{author}{\bibinfo{person}{Nima Ghorbani} {and} \bibinfo{person}{Michael~J. Black}.} \bibinfo{year}{2021}\natexlab{}.
\newblock \showarticletitle{{SOMA}: Solving Optical Marker-Based MoCap Automatically}. In \bibinfo{booktitle}{\emph{Proceedings of IEEE/CVF International Conference on Computer Vision (ICCV)}}.
\newblock


\bibitem[Ghosh et~al\mbox{.}(2023)]%
        {ghosh2023imos}
\bibfield{author}{\bibinfo{person}{Anindita Ghosh}, \bibinfo{person}{Rishabh Dabral}, \bibinfo{person}{Vladislav Golyanik}, \bibinfo{person}{Christian Theobalt}, {and} \bibinfo{person}{Philipp Slusallek}.} \bibinfo{year}{2023}\natexlab{}.
\newblock \showarticletitle{IMoS: Intent-Driven Full-Body Motion Synthesis for Human-Object Interactions}. In \bibinfo{booktitle}{\emph{Computer Graphics Forum}}, Vol.~\bibinfo{volume}{42}. Wiley Online Library, \bibinfo{pages}{1--12}.
\newblock


\bibitem[Hassan et~al\mbox{.}(2021)]%
        {hassan2021stochastic}
\bibfield{author}{\bibinfo{person}{Mohamed Hassan}, \bibinfo{person}{Duygu Ceylan}, \bibinfo{person}{Ruben Villegas}, \bibinfo{person}{Jun Saito}, \bibinfo{person}{Jimei Yang}, \bibinfo{person}{Yi Zhou}, {and} \bibinfo{person}{Michael~J Black}.} \bibinfo{year}{2021}\natexlab{}.
\newblock \showarticletitle{Stochastic scene-aware motion prediction}. In \bibinfo{booktitle}{\emph{Proceedings of the IEEE/CVF International Conference on Computer Vision}}. \bibinfo{pages}{11374--11384}.
\newblock


\bibitem[Hassan et~al\mbox{.}(2019)]%
        {hassan2019resolving}
\bibfield{author}{\bibinfo{person}{Mohamed Hassan}, \bibinfo{person}{Vasileios Choutas}, \bibinfo{person}{Dimitrios Tzionas}, {and} \bibinfo{person}{Michael~J Black}.} \bibinfo{year}{2019}\natexlab{}.
\newblock \showarticletitle{Resolving 3D human pose ambiguities with 3D scene constraints}. In \bibinfo{booktitle}{\emph{Proceedings of the IEEE/CVF international conference on computer vision}}. \bibinfo{pages}{2282--2292}.
\newblock


\bibitem[Hassan et~al\mbox{.}(2023)]%
        {hassan2023synthesizing}
\bibfield{author}{\bibinfo{person}{Mohamed Hassan}, \bibinfo{person}{Yunrong Guo}, \bibinfo{person}{Tingwu Wang}, \bibinfo{person}{Michael Black}, \bibinfo{person}{Sanja Fidler}, {and} \bibinfo{person}{Xue~Bin Peng}.} \bibinfo{year}{2023}\natexlab{}.
\newblock \showarticletitle{Synthesizing Physical Character-Scene Interactions}.
\newblock \bibinfo{journal}{\emph{SIGGRAPH Conference}} (\bibinfo{year}{2023}).
\newblock


\bibitem[Holden et~al\mbox{.}(2017)]%
        {holden2017phase}
\bibfield{author}{\bibinfo{person}{Daniel Holden}, \bibinfo{person}{Taku Komura}, {and} \bibinfo{person}{Jun Saito}.} \bibinfo{year}{2017}\natexlab{}.
\newblock \showarticletitle{Phase-functioned neural networks for character control}.
\newblock \bibinfo{journal}{\emph{ACM Transactions on Graphics (TOG)}} \bibinfo{volume}{36}, \bibinfo{number}{4} (\bibinfo{year}{2017}), \bibinfo{pages}{1--13}.
\newblock


\bibitem[Hu et~al\mbox{.}(2024)]%
        {hu2024hand}
\bibfield{author}{\bibinfo{person}{Haoyu Hu}, \bibinfo{person}{Xinyu Yi}, \bibinfo{person}{Zhe Cao}, \bibinfo{person}{Jun-Hai Yong}, {and} \bibinfo{person}{Feng Xu}.} \bibinfo{year}{2024}\natexlab{}.
\newblock \showarticletitle{Hand-Object Interaction Controller (HOIC): Deep Reinforcement Learning for Reconstructing Interactions with Physics}. In \bibinfo{booktitle}{\emph{ACM SIGGRAPH 2024 Conference Papers}}. \bibinfo{pages}{1--10}.
\newblock


\bibitem[Karunratanakul et~al\mbox{.}(2020)]%
        {karunratanakul2020grasping}
\bibfield{author}{\bibinfo{person}{Korrawe Karunratanakul}, \bibinfo{person}{Jinlong Yang}, \bibinfo{person}{Yan Zhang}, \bibinfo{person}{Michael~J Black}, \bibinfo{person}{Krikamol Muandet}, {and} \bibinfo{person}{Siyu Tang}.} \bibinfo{year}{2020}\natexlab{}.
\newblock \showarticletitle{Grasping field: Learning implicit representations for human grasps}. In \bibinfo{booktitle}{\emph{2020 International Conference on 3D Vision (3DV)}}. IEEE, \bibinfo{pages}{333--344}.
\newblock


\bibitem[Lee and Joo(2023)]%
        {lee2023lama}
\bibfield{author}{\bibinfo{person}{Jiye Lee} {and} \bibinfo{person}{Hanbyul Joo}.} \bibinfo{year}{2023}\natexlab{}.
\newblock \showarticletitle{Locomotion-Action-Manipulation: Synthesizing Human-Scene Interactions in Complex 3D Environments}.
\newblock \bibinfo{journal}{\emph{International conference on computer vision (ICCV)}} (\bibinfo{year}{2023}).
\newblock


\bibitem[Li et~al\mbox{.}(2024)]%
        {li2024controllable}
\bibfield{author}{\bibinfo{person}{Jiaman Li}, \bibinfo{person}{Alexander Clegg}, \bibinfo{person}{Roozbeh Mottaghi}, \bibinfo{person}{Jiajun Wu}, \bibinfo{person}{Xavier Puig}, {and} \bibinfo{person}{C~Karen Liu}.} \bibinfo{year}{2024}\natexlab{}.
\newblock \showarticletitle{Controllable human-object interaction synthesis}. In \bibinfo{booktitle}{\emph{European Conference on Computer Vision}}. Springer, \bibinfo{pages}{54--72}.
\newblock


\bibitem[Li et~al\mbox{.}(2023)]%
        {li2023object}
\bibfield{author}{\bibinfo{person}{Jiaman Li}, \bibinfo{person}{Jiajun Wu}, {and} \bibinfo{person}{C~Karen Liu}.} \bibinfo{year}{2023}\natexlab{}.
\newblock \showarticletitle{Object motion guided human motion synthesis}.
\newblock \bibinfo{journal}{\emph{ACM Transactions on Graphics (TOG)}} \bibinfo{volume}{42}, \bibinfo{number}{6} (\bibinfo{year}{2023}), \bibinfo{pages}{1--11}.
\newblock


\bibitem[Li et~al\mbox{.}(2021)]%
        {li2021learning}
\bibfield{author}{\bibinfo{person}{Xueting Li}, \bibinfo{person}{Shalini De~Mello}, \bibinfo{person}{Xiaolong Wang}, \bibinfo{person}{Ming-Hsuan Yang}, \bibinfo{person}{Jan Kautz}, {and} \bibinfo{person}{Sifei Liu}.} \bibinfo{year}{2021}\natexlab{}.
\newblock \showarticletitle{Learning continuous environment fields via implicit functions}.
\newblock \bibinfo{journal}{\emph{arXiv preprint arXiv:2111.13997}} (\bibinfo{year}{2021}).
\newblock


\bibitem[Loper et~al\mbox{.}(2023)]%
        {loper2023smpl}
\bibfield{author}{\bibinfo{person}{Matthew Loper}, \bibinfo{person}{Naureen Mahmood}, \bibinfo{person}{Javier Romero}, \bibinfo{person}{Gerard Pons-Moll}, {and} \bibinfo{person}{Michael~J Black}.} \bibinfo{year}{2023}\natexlab{}.
\newblock \showarticletitle{SMPL: A skinned multi-person linear model}.
\newblock In \bibinfo{booktitle}{\emph{Seminal Graphics Papers: Pushing the Boundaries, Volume 2}}. \bibinfo{pages}{851--866}.
\newblock


\bibitem[Luo et~al\mbox{.}(2024)]%
        {luo2024omnigrasp}
\bibfield{author}{\bibinfo{person}{Zhengyi Luo}, \bibinfo{person}{Jinkun Cao}, \bibinfo{person}{Sammy Christen}, \bibinfo{person}{Alexander Winkler}, \bibinfo{person}{Kris Kitani}, {and} \bibinfo{person}{Weipeng Xu}.} \bibinfo{year}{2024}\natexlab{}.
\newblock \showarticletitle{Omnigrasp: Grasping diverse objects with simulated humanoids}.
\newblock \bibinfo{journal}{\emph{Advances in Neural Information Processing Systems}}  \bibinfo{volume}{37} (\bibinfo{year}{2024}), \bibinfo{pages}{2161--2184}.
\newblock


\bibitem[Luo et~al\mbox{.}(2023)]%
        {luo2023universal}
\bibfield{author}{\bibinfo{person}{Zhengyi Luo}, \bibinfo{person}{Jinkun Cao}, \bibinfo{person}{Josh Merel}, \bibinfo{person}{Alexander Winkler}, \bibinfo{person}{Jing Huang}, \bibinfo{person}{Kris Kitani}, {and} \bibinfo{person}{Weipeng Xu}.} \bibinfo{year}{2023}\natexlab{}.
\newblock \showarticletitle{Universal humanoid motion representations for physics-based control}.
\newblock \bibinfo{journal}{\emph{arXiv preprint arXiv:2310.04582}} (\bibinfo{year}{2023}).
\newblock


\bibitem[Mahmood et~al\mbox{.}(2019)]%
        {AMASS:ICCV:2019}
\bibfield{author}{\bibinfo{person}{Naureen Mahmood}, \bibinfo{person}{Nima Ghorbani}, \bibinfo{person}{Nikolaus~F. Troje}, \bibinfo{person}{Gerard Pons-Moll}, {and} \bibinfo{person}{Michael~J. Black}.} \bibinfo{year}{2019}\natexlab{}.
\newblock \showarticletitle{{AMASS}: Archive of Motion Capture as Surface Shapes}. In \bibinfo{booktitle}{\emph{International Conference on Computer Vision}}. \bibinfo{pages}{5442--5451}.
\newblock


\bibitem[Manuelli et~al\mbox{.}(2019)]%
        {manuelli2019kpam}
\bibfield{author}{\bibinfo{person}{Lucas Manuelli}, \bibinfo{person}{Wei Gao}, \bibinfo{person}{Peter Florence}, {and} \bibinfo{person}{Russ Tedrake}.} \bibinfo{year}{2019}\natexlab{}.
\newblock \showarticletitle{KPAM: Keypoint affordances for category-level robotic manipulation}. In \bibinfo{booktitle}{\emph{The International Symposium of Robotics Research}}. Springer, \bibinfo{pages}{132--157}.
\newblock


\bibitem[Marschner et~al\mbox{.}(2023)]%
        {marschner2023constructive}
\bibfield{author}{\bibinfo{person}{Zo{\"e} Marschner}, \bibinfo{person}{Silvia Sell{\'a}n}, \bibinfo{person}{Hsueh-Ti~Derek Liu}, {and} \bibinfo{person}{Alec Jacobson}.} \bibinfo{year}{2023}\natexlab{}.
\newblock \showarticletitle{Constructive solid geometry on neural signed distance fields}. In \bibinfo{booktitle}{\emph{SIGGRAPH Asia 2023 Conference Papers}}. \bibinfo{pages}{1--12}.
\newblock


\bibitem[Merel et~al\mbox{.}(2020)]%
        {merel2020catch}
\bibfield{author}{\bibinfo{person}{Josh Merel}, \bibinfo{person}{Saran Tunyasuvunakool}, \bibinfo{person}{Arun Ahuja}, \bibinfo{person}{Yuval Tassa}, \bibinfo{person}{Leonard Hasenclever}, \bibinfo{person}{Vu Pham}, \bibinfo{person}{Tom Erez}, \bibinfo{person}{Greg Wayne}, {and} \bibinfo{person}{Nicolas Heess}.} \bibinfo{year}{2020}\natexlab{}.
\newblock \showarticletitle{Catch \& carry: reusable neural controllers for vision-guided whole-body tasks}.
\newblock \bibinfo{journal}{\emph{ACM Transactions on Graphics (TOG)}} \bibinfo{volume}{39}, \bibinfo{number}{4} (\bibinfo{year}{2020}), \bibinfo{pages}{39--1}.
\newblock


\bibitem[Mildenhall et~al\mbox{.}(2020)]%
        {mildenhall2020nerf}
\bibfield{author}{\bibinfo{person}{Ben Mildenhall}, \bibinfo{person}{Pratul~P. Srinivasan}, \bibinfo{person}{Matthew Tancik}, \bibinfo{person}{Jonathan~T. Barron}, \bibinfo{person}{Ravi Ramamoorthi}, {and} \bibinfo{person}{Ren Ng}.} \bibinfo{year}{2020}\natexlab{}.
\newblock \showarticletitle{NeRF: Representing Scenes as Neural Radiance Fields for View Synthesis}. In \bibinfo{booktitle}{\emph{ECCV}}.
\newblock


\bibitem[Min and Chai(2012)]%
        {min2012motion}
\bibfield{author}{\bibinfo{person}{Jianyuan Min} {and} \bibinfo{person}{Jinxiang Chai}.} \bibinfo{year}{2012}\natexlab{}.
\newblock \showarticletitle{Motion graphs++ a compact generative model for semantic motion analysis and synthesis}.
\newblock \bibinfo{journal}{\emph{ACM Transactions on Graphics (TOG)}} \bibinfo{volume}{31}, \bibinfo{number}{6} (\bibinfo{year}{2012}), \bibinfo{pages}{1--12}.
\newblock


\bibitem[Mukai and Kuriyama(2005)]%
        {mukai2005geostatistical}
\bibfield{author}{\bibinfo{person}{Tomohiko Mukai} {and} \bibinfo{person}{Shigeru Kuriyama}.} \bibinfo{year}{2005}\natexlab{}.
\newblock \showarticletitle{Geostatistical motion interpolation}.
\newblock In \bibinfo{booktitle}{\emph{ACM SIGGRAPH 2005 Papers}}. \bibinfo{pages}{1062--1070}.
\newblock


\bibitem[Ni and Qureshi(2022)]%
        {ni2022ntfields}
\bibfield{author}{\bibinfo{person}{Ruiqi Ni} {and} \bibinfo{person}{Ahmed~H Qureshi}.} \bibinfo{year}{2022}\natexlab{}.
\newblock \showarticletitle{Ntfields: Neural time fields for physics-informed robot motion planning}.
\newblock \bibinfo{journal}{\emph{arXiv preprint arXiv:2210.00120}} (\bibinfo{year}{2022}).
\newblock


\bibitem[Park et~al\mbox{.}(2019)]%
        {park2019deepsdf}
\bibfield{author}{\bibinfo{person}{Jeong~Joon Park}, \bibinfo{person}{Peter Florence}, \bibinfo{person}{Julian Straub}, \bibinfo{person}{Richard Newcombe}, {and} \bibinfo{person}{Steven Lovegrove}.} \bibinfo{year}{2019}\natexlab{}.
\newblock \showarticletitle{Deepsdf: Learning continuous signed distance functions for shape representation}. In \bibinfo{booktitle}{\emph{Proceedings of the IEEE/CVF conference on computer vision and pattern recognition}}. \bibinfo{pages}{165--174}.
\newblock


\bibitem[Peng et~al\mbox{.}(2022)]%
        {peng2022ase}
\bibfield{author}{\bibinfo{person}{Xue~Bin Peng}, \bibinfo{person}{Yunrong Guo}, \bibinfo{person}{Lina Halper}, \bibinfo{person}{Sergey Levine}, {and} \bibinfo{person}{Sanja Fidler}.} \bibinfo{year}{2022}\natexlab{}.
\newblock \showarticletitle{Ase: Large-scale reusable adversarial skill embeddings for physically simulated characters}.
\newblock \bibinfo{journal}{\emph{ACM Transactions On Graphics (TOG)}} \bibinfo{volume}{41}, \bibinfo{number}{4} (\bibinfo{year}{2022}), \bibinfo{pages}{1--17}.
\newblock


\bibitem[Peng et~al\mbox{.}(2021)]%
        {peng2021amp}
\bibfield{author}{\bibinfo{person}{Xue~Bin Peng}, \bibinfo{person}{Ze Ma}, \bibinfo{person}{Pieter Abbeel}, \bibinfo{person}{Sergey Levine}, {and} \bibinfo{person}{Angjoo Kanazawa}.} \bibinfo{year}{2021}\natexlab{}.
\newblock \showarticletitle{Amp: Adversarial motion priors for stylized physics-based character control}.
\newblock \bibinfo{journal}{\emph{ACM Transactions on Graphics (ToG)}} \bibinfo{volume}{40}, \bibinfo{number}{4} (\bibinfo{year}{2021}), \bibinfo{pages}{1--20}.
\newblock


\bibitem[Pi et~al\mbox{.}(2023)]%
        {pi2023hierarchical}
\bibfield{author}{\bibinfo{person}{Huaijin Pi}, \bibinfo{person}{Sida Peng}, \bibinfo{person}{Minghui Yang}, \bibinfo{person}{Xiaowei Zhou}, {and} \bibinfo{person}{Hujun Bao}.} \bibinfo{year}{2023}\natexlab{}.
\newblock \showarticletitle{Hierarchical generation of human-object interactions with diffusion probabilistic models}. In \bibinfo{booktitle}{\emph{Proceedings of the IEEE/CVF International Conference on Computer Vision}}. \bibinfo{pages}{15061--15073}.
\newblock


\bibitem[Rempe et~al\mbox{.}(2021)]%
        {rempe2021humor}
\bibfield{author}{\bibinfo{person}{Davis Rempe}, \bibinfo{person}{Tolga Birdal}, \bibinfo{person}{Aaron Hertzmann}, \bibinfo{person}{Jimei Yang}, \bibinfo{person}{Srinath Sridhar}, {and} \bibinfo{person}{Leonidas~J Guibas}.} \bibinfo{year}{2021}\natexlab{}.
\newblock \showarticletitle{Humor: 3d human motion model for robust pose estimation}. In \bibinfo{booktitle}{\emph{Proceedings of the IEEE/CVF international conference on computer vision}}. \bibinfo{pages}{11488--11499}.
\newblock


\bibitem[Sethian(1996)]%
        {sethian1996fast}
\bibfield{author}{\bibinfo{person}{James~A Sethian}.} \bibinfo{year}{1996}\natexlab{}.
\newblock \showarticletitle{A fast marching level set method for monotonically advancing fronts.}
\newblock \bibinfo{journal}{\emph{proceedings of the National Academy of Sciences}} \bibinfo{volume}{93}, \bibinfo{number}{4} (\bibinfo{year}{1996}), \bibinfo{pages}{1591--1595}.
\newblock


\bibitem[Simeonov et~al\mbox{.}(2022)]%
        {simeonov2022neural}
\bibfield{author}{\bibinfo{person}{Anthony Simeonov}, \bibinfo{person}{Yilun Du}, \bibinfo{person}{Andrea Tagliasacchi}, \bibinfo{person}{Joshua~B Tenenbaum}, \bibinfo{person}{Alberto Rodriguez}, \bibinfo{person}{Pulkit Agrawal}, {and} \bibinfo{person}{Vincent Sitzmann}.} \bibinfo{year}{2022}\natexlab{}.
\newblock \showarticletitle{Neural descriptor fields: Se (3)-equivariant object representations for manipulation}. In \bibinfo{booktitle}{\emph{2022 International Conference on Robotics and Automation (ICRA)}}. IEEE, \bibinfo{pages}{6394--6400}.
\newblock


\bibitem[Starke et~al\mbox{.}(2022)]%
        {starke2022deepphase}
\bibfield{author}{\bibinfo{person}{Sebastian Starke}, \bibinfo{person}{Ian Mason}, {and} \bibinfo{person}{Taku Komura}.} \bibinfo{year}{2022}\natexlab{}.
\newblock \showarticletitle{Deepphase: Periodic autoencoders for learning motion phase manifolds}.
\newblock \bibinfo{journal}{\emph{ACM Transactions on Graphics (TOG)}} \bibinfo{volume}{41}, \bibinfo{number}{4} (\bibinfo{year}{2022}), \bibinfo{pages}{1--13}.
\newblock


\bibitem[Starke et~al\mbox{.}(2019)]%
        {starke2019neural}
\bibfield{author}{\bibinfo{person}{Sebastian Starke}, \bibinfo{person}{He Zhang}, \bibinfo{person}{Taku Komura}, {and} \bibinfo{person}{Jun Saito}.} \bibinfo{year}{2019}\natexlab{}.
\newblock \showarticletitle{Neural state machine for character-scene interactions.}
\newblock \bibinfo{journal}{\emph{ACM Trans. Graph.}} \bibinfo{volume}{38}, \bibinfo{number}{6} (\bibinfo{year}{2019}), \bibinfo{pages}{209--1}.
\newblock


\bibitem[Starke et~al\mbox{.}(2021)]%
        {starke2021neural}
\bibfield{author}{\bibinfo{person}{Sebastian Starke}, \bibinfo{person}{Yiwei Zhao}, \bibinfo{person}{Fabio Zinno}, {and} \bibinfo{person}{Taku Komura}.} \bibinfo{year}{2021}\natexlab{}.
\newblock \showarticletitle{Neural animation layering for synthesizing martial arts movements}.
\newblock \bibinfo{journal}{\emph{ACM Transactions on Graphics (TOG)}} \bibinfo{volume}{40}, \bibinfo{number}{4} (\bibinfo{year}{2021}), \bibinfo{pages}{1--16}.
\newblock


\bibitem[Sundaralingam et~al\mbox{.}(2023)]%
        {sundaralingam2023curobo}
\bibfield{author}{\bibinfo{person}{Balakumar Sundaralingam}, \bibinfo{person}{Siva Kumar~Sastry Hari}, \bibinfo{person}{Adam Fishman}, \bibinfo{person}{Caelan Garrett}, \bibinfo{person}{Karl Van~Wyk}, \bibinfo{person}{Valts Blukis}, \bibinfo{person}{Alexander Millane}, \bibinfo{person}{Helen Oleynikova}, \bibinfo{person}{Ankur Handa}, \bibinfo{person}{Fabio Ramos}, {et~al\mbox{.}}} \bibinfo{year}{2023}\natexlab{}.
\newblock \showarticletitle{Curobo: Parallelized collision-free robot motion generation}. In \bibinfo{booktitle}{\emph{2023 IEEE International Conference on Robotics and Automation (ICRA)}}. IEEE, \bibinfo{pages}{8112--8119}.
\newblock


\bibitem[Taheri et~al\mbox{.}(2022)]%
        {taheri2022goal}
\bibfield{author}{\bibinfo{person}{Omid Taheri}, \bibinfo{person}{Vasileios Choutas}, \bibinfo{person}{Michael~J Black}, {and} \bibinfo{person}{Dimitrios Tzionas}.} \bibinfo{year}{2022}\natexlab{}.
\newblock \showarticletitle{GOAL: Generating 4D whole-body motion for hand-object grasping}. In \bibinfo{booktitle}{\emph{Proceedings of the IEEE/CVF Conference on Computer Vision and Pattern Recognition}}. \bibinfo{pages}{13263--13273}.
\newblock


\bibitem[Taheri et~al\mbox{.}(2020)]%
        {taheri2020grab}
\bibfield{author}{\bibinfo{person}{Omid Taheri}, \bibinfo{person}{Nima Ghorbani}, \bibinfo{person}{Michael~J Black}, {and} \bibinfo{person}{Dimitrios Tzionas}.} \bibinfo{year}{2020}\natexlab{}.
\newblock \showarticletitle{GRAB: A dataset of whole-body human grasping of objects}. In \bibinfo{booktitle}{\emph{Computer Vision--ECCV 2020: 16th European Conference, Glasgow, UK, August 23--28, 2020, Proceedings, Part IV 16}}. Springer, \bibinfo{pages}{581--600}.
\newblock


\bibitem[Taheri et~al\mbox{.}(2024)]%
        {taheri2024grip}
\bibfield{author}{\bibinfo{person}{Omid Taheri}, \bibinfo{person}{Yi Zhou}, \bibinfo{person}{Dimitrios Tzionas}, \bibinfo{person}{Yang Zhou}, \bibinfo{person}{Duygu Ceylan}, \bibinfo{person}{Soren Pirk}, {and} \bibinfo{person}{Michael~J. Black}.} \bibinfo{year}{2024}\natexlab{}.
\newblock \showarticletitle{{GRIP}: Generating Interaction Poses Using Latent Consistency and Spatial Cues}. In \bibinfo{booktitle}{\emph{International Conference on 3D Vision ({3DV})}}.
\newblock
\urldef\tempurl%
\url{https://grip.is.tue.mpg.de}
\showURL{%
\tempurl}


\bibitem[Tessler et~al\mbox{.}(2024)]%
        {tessler2024maskedmimic}
\bibfield{author}{\bibinfo{person}{Chen Tessler}, \bibinfo{person}{Yunrong Guo}, \bibinfo{person}{Ofir Nabati}, \bibinfo{person}{Gal Chechik}, {and} \bibinfo{person}{Xue~Bin Peng}.} \bibinfo{year}{2024}\natexlab{}.
\newblock \showarticletitle{Maskedmimic: Unified physics-based character control through masked motion inpainting}.
\newblock \bibinfo{journal}{\emph{ACM Transactions on Graphics (TOG)}} \bibinfo{volume}{43}, \bibinfo{number}{6} (\bibinfo{year}{2024}), \bibinfo{pages}{1--21}.
\newblock


\bibitem[Tsitsiklis(1995)]%
        {tsitsiklis1995efficient}
\bibfield{author}{\bibinfo{person}{John~N Tsitsiklis}.} \bibinfo{year}{1995}\natexlab{}.
\newblock \showarticletitle{Efficient algorithms for globally optimal trajectories}.
\newblock \bibinfo{journal}{\emph{IEEE transactions on Automatic Control}} \bibinfo{volume}{40}, \bibinfo{number}{9} (\bibinfo{year}{1995}), \bibinfo{pages}{1528--1538}.
\newblock


\bibitem[Wang et~al\mbox{.}(2024)]%
        {wang2024implicit}
\bibfield{author}{\bibinfo{person}{Jingping Wang}, \bibinfo{person}{Tingrui Zhang}, \bibinfo{person}{Qixuan Zhang}, \bibinfo{person}{Chuxiao Zeng}, \bibinfo{person}{Jingyi Yu}, \bibinfo{person}{Chao Xu}, \bibinfo{person}{Lan Xu}, {and} \bibinfo{person}{Fei Gao}.} \bibinfo{year}{2024}\natexlab{}.
\newblock \showarticletitle{Implicit Swept Volume SDF: Enabling Continuous Collision-Free Trajectory Generation for Arbitrary Shapes}.
\newblock \bibinfo{journal}{\emph{ACM Transactions on Graphics (TOG)}} \bibinfo{volume}{43}, \bibinfo{number}{4} (\bibinfo{year}{2024}), \bibinfo{pages}{1--14}.
\newblock


\bibitem[Wang et~al\mbox{.}(2022)]%
        {wang2022geometrically}
\bibfield{author}{\bibinfo{person}{Zhepei Wang}, \bibinfo{person}{Xin Zhou}, \bibinfo{person}{Chao Xu}, {and} \bibinfo{person}{Fei Gao}.} \bibinfo{year}{2022}\natexlab{}.
\newblock \showarticletitle{Geometrically constrained trajectory optimization for multicopters}.
\newblock \bibinfo{journal}{\emph{IEEE Transactions on Robotics}} \bibinfo{volume}{38}, \bibinfo{number}{5} (\bibinfo{year}{2022}), \bibinfo{pages}{3259--3278}.
\newblock


\bibitem[Wu et~al\mbox{.}(2022)]%
        {wu2022saga}
\bibfield{author}{\bibinfo{person}{Yan Wu}, \bibinfo{person}{Jiahao Wang}, \bibinfo{person}{Yan Zhang}, \bibinfo{person}{Siwei Zhang}, \bibinfo{person}{Otmar Hilliges}, \bibinfo{person}{Fisher Yu}, {and} \bibinfo{person}{Siyu Tang}.} \bibinfo{year}{2022}\natexlab{}.
\newblock \showarticletitle{Saga: Stochastic whole-body grasping with contact}. In \bibinfo{booktitle}{\emph{European Conference on Computer Vision}}. Springer, \bibinfo{pages}{257--274}.
\newblock


\bibitem[Xu et~al\mbox{.}(2023)]%
        {xu2023interdiff}
\bibfield{author}{\bibinfo{person}{Sirui Xu}, \bibinfo{person}{Zhengyuan Li}, \bibinfo{person}{Yu-Xiong Wang}, {and} \bibinfo{person}{Liang-Yan Gui}.} \bibinfo{year}{2023}\natexlab{}.
\newblock \showarticletitle{InterDiff: Generating 3D Human-Object Interactions with Physics-Informed Diffusion}. In \bibinfo{booktitle}{\emph{International conference on computer vision (ICCV)}}.
\newblock


\bibitem[Yang et~al\mbox{.}(2022)]%
        {yang2022oakink}
\bibfield{author}{\bibinfo{person}{Lixin Yang}, \bibinfo{person}{Kailin Li}, \bibinfo{person}{Xinyu Zhan}, \bibinfo{person}{Fei Wu}, \bibinfo{person}{Anran Xu}, \bibinfo{person}{Liu Liu}, {and} \bibinfo{person}{Cewu Lu}.} \bibinfo{year}{2022}\natexlab{}.
\newblock \showarticletitle{OakInk: A large-scale knowledge repository for understanding hand-object interaction}. In \bibinfo{booktitle}{\emph{Proceedings of the IEEE/CVF Conference on Computer Vision and Pattern Recognition}}. \bibinfo{pages}{20953--20962}.
\newblock


\bibitem[Ye and Liu(2012)]%
        {ye2012synthesis}
\bibfield{author}{\bibinfo{person}{Yuting Ye} {and} \bibinfo{person}{C~Karen Liu}.} \bibinfo{year}{2012}\natexlab{}.
\newblock \showarticletitle{Synthesis of detailed hand manipulations using contact sampling}.
\newblock \bibinfo{journal}{\emph{ACM Transactions on Graphics (ToG)}} \bibinfo{volume}{31}, \bibinfo{number}{4} (\bibinfo{year}{2012}), \bibinfo{pages}{1--10}.
\newblock


\bibitem[Zhang et~al\mbox{.}(2024)]%
        {zhang2024artigrasp}
\bibfield{author}{\bibinfo{person}{Hui Zhang}, \bibinfo{person}{Sammy Christen}, \bibinfo{person}{Zicong Fan}, \bibinfo{person}{Luocheng Zheng}, \bibinfo{person}{Jemin Hwangbo}, \bibinfo{person}{Jie Song}, {and} \bibinfo{person}{Otmar Hilliges}.} \bibinfo{year}{2024}\natexlab{}.
\newblock \showarticletitle{{ArtiGrasp}: Physically Plausible Synthesis of Bi-Manual Dexterous Grasping and Articulation}. In \bibinfo{booktitle}{\emph{International Conference on 3D Vision (3DV)}}.
\newblock


\bibitem[Zhang et~al\mbox{.}(2018)]%
        {zhang2018mode}
\bibfield{author}{\bibinfo{person}{He Zhang}, \bibinfo{person}{Sebastian Starke}, \bibinfo{person}{Taku Komura}, {and} \bibinfo{person}{Jun Saito}.} \bibinfo{year}{2018}\natexlab{}.
\newblock \showarticletitle{Mode-adaptive neural networks for quadruped motion control}.
\newblock \bibinfo{journal}{\emph{ACM Transactions on Graphics (TOG)}} \bibinfo{volume}{37}, \bibinfo{number}{4} (\bibinfo{year}{2018}), \bibinfo{pages}{1--11}.
\newblock


\bibitem[Zhang et~al\mbox{.}(2021)]%
        {zhang2021manipnet}
\bibfield{author}{\bibinfo{person}{He Zhang}, \bibinfo{person}{Yuting Ye}, \bibinfo{person}{Takaaki Shiratori}, {and} \bibinfo{person}{Taku Komura}.} \bibinfo{year}{2021}\natexlab{}.
\newblock \showarticletitle{Manipnet: neural manipulation synthesis with a hand-object spatial representation}.
\newblock \bibinfo{journal}{\emph{ACM Transactions on Graphics (ToG)}} \bibinfo{volume}{40}, \bibinfo{number}{4} (\bibinfo{year}{2021}), \bibinfo{pages}{1--14}.
\newblock


\bibitem[Zhang et~al\mbox{.}(2022)]%
        {zhang2022couch}
\bibfield{author}{\bibinfo{person}{Xiaohan Zhang}, \bibinfo{person}{Bharat~Lal Bhatnagar}, \bibinfo{person}{Sebastian Starke}, \bibinfo{person}{Vladimir Guzov}, {and} \bibinfo{person}{Gerard Pons-Moll}.} \bibinfo{year}{2022}\natexlab{}.
\newblock \showarticletitle{COUCH: Towards Controllable Human-Chair Interactions}.
\newblock  (\bibinfo{date}{October} \bibinfo{year}{2022}).
\newblock


\bibitem[Zhao et~al\mbox{.}(2022)]%
        {Zhao:ECCV:2022}
\bibfield{author}{\bibinfo{person}{Kaifeng Zhao}, \bibinfo{person}{Shaofei Wang}, \bibinfo{person}{Yan Zhang}, \bibinfo{person}{Thabo Beeler}, \bibinfo{person}{}, {and} \bibinfo{person}{Siyu Tang}.} \bibinfo{year}{2022}\natexlab{}.
\newblock \showarticletitle{Compositional Human-Scene Interaction Synthesis with Semantic Control}. In \bibinfo{booktitle}{\emph{European conference on computer vision (ECCV)}}.
\newblock


\bibitem[Zhao et~al\mbox{.}(2023a)]%
        {Zhao:ICCV:2023}
\bibfield{author}{\bibinfo{person}{Kaifeng Zhao}, \bibinfo{person}{Yan Zhang}, \bibinfo{person}{Shaofei Wang}, \bibinfo{person}{Thabo Beeler}, {and} \bibinfo{person}{Siyu Tang}.} \bibinfo{year}{2023}\natexlab{a}.
\newblock \showarticletitle{DIMOS: Synthesizing Diverse Human Motions in 3D Indoor Scenes}. In \bibinfo{booktitle}{\emph{International conference on computer vision (ICCV)}}.
\newblock


\bibitem[Zhao et~al\mbox{.}(2023b)]%
        {zhao2023synthesizing}
\bibfield{author}{\bibinfo{person}{Kaifeng Zhao}, \bibinfo{person}{Yan Zhang}, \bibinfo{person}{Shaofei Wang}, \bibinfo{person}{Thabo Beeler}, {and} \bibinfo{person}{Siyu Tang}.} \bibinfo{year}{2023}\natexlab{b}.
\newblock \showarticletitle{Synthesizing diverse human motions in 3d indoor scenes}. In \bibinfo{booktitle}{\emph{Proceedings of the IEEE/CVF International Conference on Computer Vision}}. \bibinfo{pages}{14738--14749}.
\newblock


\bibitem[Zheng et~al\mbox{.}(2023)]%
        {Zheng_2023_CVPR}
\bibfield{author}{\bibinfo{person}{Juntian Zheng}, \bibinfo{person}{Qingyuan Zheng}, \bibinfo{person}{Lixing Fang}, \bibinfo{person}{Yun Liu}, {and} \bibinfo{person}{Li Yi}.} \bibinfo{year}{2023}\natexlab{}.
\newblock \showarticletitle{CAMS: CAnonicalized Manipulation Spaces for Category-Level Functional Hand-Object Manipulation Synthesis}. In \bibinfo{booktitle}{\emph{Proceedings of the IEEE/CVF Conference on Computer Vision and Pattern Recognition (CVPR)}}. \bibinfo{pages}{585--594}.
\newblock


\bibitem[Zhou et~al\mbox{.}(2019)]%
        {zhou2019robust}
\bibfield{author}{\bibinfo{person}{Boyu Zhou}, \bibinfo{person}{Fei Gao}, \bibinfo{person}{Luqi Wang}, \bibinfo{person}{Chuhao Liu}, {and} \bibinfo{person}{Shaojie Shen}.} \bibinfo{year}{2019}\natexlab{}.
\newblock \showarticletitle{Robust and efficient quadrotor trajectory generation for fast autonomous flight}.
\newblock \bibinfo{journal}{\emph{IEEE Robotics and Automation Letters}} \bibinfo{volume}{4}, \bibinfo{number}{4} (\bibinfo{year}{2019}), \bibinfo{pages}{3529--3536}.
\newblock


\end{thebibliography}
